 \documentclass[preprint,showpacs,preprintnumbers,amsmath,amssymb, superscriptaddress,aps,prd,longbibliography,nofootinbib]{revtex4-1}
\usepackage{graphicx}
\usepackage{xcolor}
\usepackage{amsmath}
\usepackage{array}
\usepackage{hyperref}
\usepackage{subfigure}
\usepackage{amsfonts}
\usepackage[top = 2cm, bottom = 2cm, right = 2cm, left =2cm]{geometry}
\usepackage{amssymb}
\graphicspath{ {figs/} }
\usepackage[latin1]{inputenc}
\usepackage[english]{babel}
\usepackage{epsfig}
\usepackage{wasysym}
\usepackage{color,xcolor}

\begin{document}

\title{Field Sources for $f(R,R_{\mu\nu})$ Black-Bounce Solutions: The Case of K-Gravity}

\author{G. Alencar}
\email{geova@fisica.ufc.br}
\affiliation{Departamento de F\'isica, Universidade Federal do Cear\'a, Caixa Postal 6030, Campus do Pici, 60455-760 Fortaleza, Cear\'a, Brazil}

\author{M. Nilton}
\email{matheus.nilton@fisica.ufc.br}
\affiliation{Universidade Federal do Cear\'{a} Campus de Russas, 62900-420, Russas, Cear\'{a}, Brazil}

 \author{Manuel E. Rodrigues} 
 \email{esialg@gmail.com}
\affiliation{Faculdade de F\'{i}sica, Programa de P\'{o}s-Gradua\c{c}\~{a}o em F\'{i}sica, Universidade Federal do Par\'{a}, 66075-110, Bel\'{e}m, Par\'{a}, Brazill}
\affiliation{Faculdade de Ci\^{e}ncias Exatas e Tecnologia, Universidade Federal do Par\'{a}, Campus Universit\'{a}rio de Abaetetuba, 68440-000, Abaetetuba, Par\'{a}, Brazil}

\author{Marcos V. de S. Silva}
\email{marco2s303@gmail.com}
\affiliation{Departamento de F\'isica, Universidade Federal do Cear\'a, Caixa Postal 6030, Campus do Pici, 60455-760 Fortaleza, Cear\'a, Brazil}

\date{\today}

\begin{abstract}
In the framework of Simpson-Visser, the search for field sources that produce black-bounces in alternative gravity theories has remained unresolved. In this paper, the first in a series exploring sources for alternative theories of gravity, we identify such a source for the $2+1$ dimensional K-gravity black-bounce. The K-gravity black hole is notable for allowing asymptotically locally flat solutions in lower-dimensional spacetime, yet it possesses curvature singularities concealed within the event horizon. Using the Simpson-Visser regularization technique, we eliminate this singularity, constructing asymptotically locally flat black-bounce solutions in $2+1$ dimensions. We explore the causal structure of these solutions, identifying the conditions under which they describe regular black holes or wormholes. By calculating curvature invariants, we confirm the absence of singularities within the event horizon. Additionally, we demonstrate that, beyond non-linear electrodynamics, a non-linear scalar field is required to source the solution. Finally, we investigate the geodesic structure of this spacetime, analyzing the trajectories of both massive and massless particles. We also confirm the existence of circular orbits and assess their stability.

\end{abstract}

\maketitle

\section{Introduction}\label{S:Introduction}
General Relativity (GR) has garnered significant interest in recent years, driven by advancements in high-precision measurements of phenomena like black holes and cosmology. This period has witnessed milestones such as the direct detection of gravitational waves by LIGO and VIRGO, the image of supermassive black holes by the Event Horizon Telescope (EHT), and precise assessments of the $\Lambda CDM$ parameters \cite{LIGOScientific:2016aoc, EventHorizonTelescope:2022wkp, EventHorizonTelescope:2019dse, LIGOScientific:2017vwq, Planck:2018vyg, ACT:2020gnv, SDSS:2006srq}. However, the interior of black holes presents a challenge due to spacetime singularities.

A valuable avenue for probing black hole characteristics is through its $2+1$ dimensional counterpart, introduced in 1992 by Banados, Teitelboim, and Zanelli \cite{Banados:1992wn}. In the $2+1$ Einstein-Hilbert theory, the classical level presents a trivial scenario devoid of gravitational waves, where any two solutions are locally equivalent. Interestingly, the $2+1$ dimensional vacuum lacks local degrees of freedom \cite{Leutwyler1966AM}, making black holes unexpected since a vacuum solution in $2+1$ dimensions is typically flat. However, by introducing a cosmological constant with $\Lambda < 0$, Banados, Teitelboim, and Zanelli uncovered the renowned BTZ black holes \cite{Banados:1992wn}.

In the context of GR, it is possible to avoid the singularity problem through the models known as regular black hole \cite{Ansoldi:2008jw} (RBH). These spacetimes do not present singularities inside them, in such a way that the geodesics are not interrupted \cite{Bronnikov:2022ofk}. The first model of a RBH was proposed by Bardeen and was not, until then, a solution to the field equations of gravitational theories. Beato and Garcia demonstrated that the Bardeen solution could be obtained in GR if a non-linear electrodynamics (NED) were considered as the source \cite{Ayon-Beato:2000mjt,Rodrigues:2018bdc}. These black holes, which undergo a regularization process that allows $r=0$ to be a regular point, similar to Bardeen's model, are extensively studied in the literature \cite{Culetu:2014lca,Balart:2014cga,Dymnikova:2004zc,Bronnikov:2000vy,Ayon-Beato:1999kuh,Rodrigues:2020pem,Macedo:2024dqb,Paul:2023pqn}. They violate the strong energy condition in some region within the event horizon \cite{Rodrigues:2017yry} and can be derived from the coupling between GR and a NED, at least in static cases \cite{Bronnikov:2022ofk}.

In the context of modified theories of gravity, it is possible to find regular black hole solutions without the need for exotic fields. In particular, in \cite{Bueno:2024dgm}, the authors show that it is possible to construct regular solutions in vacuum by considering an infinite number of high-curvature corrections to the Einstein-Hilbert action. These corrections do not have connections with the matter sector, thus not requiring additional fields to regularize such spacetimes. In a different formulation, Konoplya and Zhidenko showed that it is also possible to obtain a $D$-dimensional generalization of Dymnikova's regular black hole as a vacuum solution \cite{Konoplya:2024kih}. Since then, other solutions in the context of pure gravity have been proposed in the literature, and various properties of these solutions have been studied \cite{Konoplya:2024hfg,DiFilippo:2024mwm,Giacchini:2024exc,Wang:2024zlq,Ditta:2024iky}.

In another direction, Simpson and Visser proposed a different method to avoid the singularity problem \cite{Simpson_2019} by introducing a regularization parameter $a^2$. Depending on the value of $a$, the four-dimensional spacetime could represent a traversable wormhole (TWH), a one-way wormhole (OWWH) with an extremal null throat at $r=0$, or a RBH. This type of solution is known as a black-bounce (BB). This method has been recently used to regularize cylindrical black holes, such as BTZ and $4d$ black strings \cite{Furtado:2022tnb, Lima:2022pvc, Lima:2023arg, Lima:2023jtl, Alencar:2024yvh, Bronnikov:2023aya}.  There are also some models of BB that are not derived from the Simpson-Visser regularization method \cite{Huang:2019arj, Lobo:2020ffi, Rodrigues:2022mdm}.

An important point in the solutions of BB, within the context of GR, is that for this type of spacetime to be a solution to Einstein equations, it is necessary to couple the gravitational theory with a NED and a phantom scalar field. Neither the scalar field alone nor the NED by itself are capable of generating these objects in GR \cite{Canate:2022gpy,Bronnikov:2021uta,Rodrigues:2023vtm,Bronnikov:2023aya,Lima:2023arg,Alencar:2024yvh,Rodrigues:2025plw,Pereira:2024gsl,Pereira:2024rtv}. In some cases, in alternative theories of gravity, it becomes simpler to consider anisotropic fluids rather than trying to determine the specific source of these fluids \cite{Fabris:2023opv,Junior:2023qaq,Atazadeh:2023wdw,Rodrigues:2024iao}.

Recently, Bronnikov proposed a new regularization method. In this method, the singular region of spacetime is not replaced by a wormhole at its center. Instead, the solution is regularized by simply `curing' the singularity \cite{Bronnikov:2024izh,Bolokhov:2024sdy}. Depending on how the `cure' is applied, the source of these solutions can be either NED alone or a combination of NED and a scalar field.

A prominent three dimensional GR modification is known as ``topologically massive gravity'' (TMG). This theory extends the Einstein-Hilbert (EH) action (plus the cosmological constant) by incorporating a Lorentz-Chern-Simons (LCS) term \cite{Deser:1981wh}, leading to the following equation:
\begin{equation}
    \sigma G_{\mu\nu}+\frac{1}{\mu}C_{\mu\nu}+\Lambda_0 g_{\mu\nu}=0,
\end{equation}
where $\sigma$ is a sign (plus for GR and minus for TMG). The TMG violates parity, however, recently, a novel formulation of massive gravity that is parity preserving was proposed by Bergshoeff, Hohm and Townsend (BHT). They added to the EH action a quadratic term, to obtain
\begin{equation}\label{ouraction}
 S_{BHT}= \frac{1}{\kappa^{2}}\int \! d^3 x\, \left\{ \sqrt{|g|} \left[ R  - \frac{1}{m^{2}} K
-2\lambda \right]\right\}\, ,
\end{equation}
where $g$ is the determinant of the metric $g_{\mu\nu}$, $m$ is a `relative' mass parameter, which could be traded for the effective dimensionless coupling constant $m \kappa^2$, and $K =  R_{\mu\nu}R^{\mu\nu} -\frac{3}{8} R^2$. 

This New Massive Gravity (NMG) extends Einstein-Hilbert gravity in three dimensions, introducing a massive graviton without ghost instabilities, making it relevant for quantum gravity, holography, and modified gravity models \cite{Bergshoeff:2009hq}. Unlike four-dimensional massive gravity, NMG avoids the Boulware Deser ghost \cite{Boulware:1972yco} and preserves unitarity by maintaining only a massive spin-2 mode \cite{Deser:2009hb,Paulos:2010ke}. It features a two-helicity massive graviton similar to Fierz-Pauli gravity in four dimensions, ensuring a consistent spin-2 excitation with two dynamical degrees of freedom \cite{Bergshoeff:2009aq,deRham:2014zqa}. Studies on AdS backgrounds confirm its relevance in holographic applications, particularly in the AdS/CFT correspondence and holographic renormalisation \cite{Alishahiha:2010bw,Parvizi:2017boc}. NMG also provides a viable framework for extending GR in three dimensions \cite{Myung:2012ee} and exploring bi-gravity models where massive and massless spin-2 fields interact \cite{Banados:2009it}. Furthermore, its incorporation into supergravity suggests potential quantum completions of massive gravity \cite{Bergshoeff:2009hq}.

Soon later, a unique maximally symmetric solution was found, given by \cite{Bergshoeff:2009aq}
\begin{equation}
    ds^2=f(r)dt^2-f(r)^{-1}dr^2-r^2d\varphi^2, \quad f(r)=-\Lambda r^2+br-\mu,
\end{equation}
where $\Lambda=2\lambda$ and $b$ and $\mu$ are constants. With $b = 0$, the solution corresponds to the static BTZ black hole with a twice mass parameter. The authors studied the AdS solution with $\Lambda=-1/l^2$.  However, in Ref. \cite{Oliva:2009ip}, an asymptotically locally flat\footnote{The term `asymptotically locally flat' would be an appropriate way to describe this type of metric. The idea behind this term is to capture the fact that spacetime tends to be locally flat at infinity (with the curvature invariants approaching zero) without necessarily tending to the Minkowski metric globally.} black hole solution was found, with $\Lambda=0$. This is remarkable since it is generally believed that only AdS solutions are allowed in three dimensions. Further properties were studied in Refs. \cite{Barnich:2015dvt, Alkac:2016xlr}.

For the asymptotically locally flat solution, with $\Lambda=0$, we have
\begin{equation}
    f(r)=b r-\mu.
\end{equation}
We see that we have a black hole horizon only in the case $\mu>0$. Computing the mass parameter, the curvature scalar and the Kretschmann scalar,\footnote{The Kretschmann scalar is defined as $\mathcal{K}=R^{\mu\nu\alpha\beta}R_{\mu\nu\alpha\beta}$, where $R^{\mu\nu\alpha\beta}$ is the curvature tensor.} we find
\begin{equation}
    M=\frac{1+\mu}{4G},\;\quad R=\frac{2b}{r}, \;\quad \mathcal{K}=\frac{2b^2}{r^2}.
\end{equation}
Although the solution does not directly recover Minkowski at infinity, by the form of these scalars, we can see that the solution is asymptotically locally flat and has a singularity at $r=0$. The singularity is hidden by the presence of the event horizon, located at $r_h=\mu/b$.

For $\Lambda=-1/l^2$, we have AdS solutions, with mass parameter and Ricci scalar given by
\begin{equation}\label{horizonBHT}
    r_{\pm}=\frac{l^2}{2}\left(-b\pm\sqrt{b^2+\frac{4\mu}{l^2}}\right), \; M=\frac{1+\mu}{4G}, \;  R=\frac{6}{l^2}+\frac{2b}{r}, \;\mathcal{K}=\frac{4}{l^4}+\frac{2}{r^2}\left(b+\frac{2r}{l^2}\right)^2.
\end{equation}
We must have $\mu\geq -\frac{1}{4}b^2l^2$ and since the mass is solely determined by $\mu$, the constant $b$ can be interpreted as a type of "gravitational hair." This is further supported by the fact that the black hole does not possess any additional global charges arising from asymptotic symmetries \cite{Oliva:2009ip}. In contrast, introducing the gravitational hair parameter $b$ causes a curvature singularity to form at the origin, as illustrated in equation \eqref{horizonBHT}. We point out that (\ref{horizonBHT}) is very similar to the Reissner-Nordstrom(RN) case and a Cauchy horizon can appear. However, now, $b$ and $\mu$ can have negative ranges which results in a naked singularity, one or two horizons. We describe in detail below:
\begin{enumerate}
    \item $b > 0$: For $\mu < 0$, we have no horizon and a naked singularity. For $\mu \ge 0$, we have a single event horizon located at $r = r_+$.
  \begin{itemize}
      \item For $\mu > 0$, the event horizon surrounds a spacelike singularity.
      \item For $\mu = 0$, the horizon coincides with a null singularity.
  \end{itemize} 
  \item $b < 0$ : The singularity is surrounded by an event horizon, and there is also the possibility to a Cauchy horizon.
\begin{itemize}
    \item For $\mu > 0$, there is a single event horizon at $r = r_+$, enclosing a spacelike singularity.
\item For $\mu = 0$, we have an event horizon, and an inner horizon appears on top of a null singularity.
\item For $-\frac{1}{4}b^2l^2< \mu < 0$, we have an inner Cauchy horizon at $r=r_-$, surrounding a timelike singularity, and an event horizon located at $r = r_+$.
\item For $\mu= -b^2l^2/4$, both horizons
coincide, enclosing a timelike singularity.
\end{itemize}
\end{enumerate}
This way, we see that asymptotically locally flat solutions in $2+1$ dimensions are possible and hold significant theoretical interest. Therefore, studying regularized versions of these solutions becomes very interesting as they no longer exhibit pathologies such as naked singularities.

Despite the notable features of $K$--gravity, it represents a specific instance within a much broader class of gravitational theories known as $f(R,R_{\mu\nu})$ theories. In these more general frameworks, the gravitational action is extended beyond the standard Einstein-Hilbert action by allowing the Lagrangian to be an arbitrary function not only of the Ricci scalar $R$ but also of invariants constructed from the Ricci tensor (e.g. $R_{\mu\nu}R^{\mu\nu}$). For example, Stelle \cite{Stelle:1976gc} demonstrated that including quadratic curvature corrections can lead to a renormalizable theory of gravity, although such theories typically suffer from the appearance of ghost modes. Whitt \cite{Whitt:1984pd} later showed that fourth-order gravity models can be reformulated as GR coupled to additional scalar fields, thus providing an alternative interpretation of the extra degrees of freedom. Numerous studies exploring this class of theories have been carried out in the literature, encompassing both cosmological applications and other physical scenarios \cite{Olmo:2012vd,Li:2008fa,Olmo:2009xy}. In this context, the $K$--gravity model is particularly compelling since it naturally emerges as a specific case in the $f(R,R_{\mu\nu})$ family, offering valuable insights into regularizing black hole singularities and exploring modified gravitational dynamics purely in the vacuum.

However, up to now, field sources for BB of alternative theories of gravity, such as $f(R,R_{\mu\nu})$, have never been found. Therefore, this is the first in a series of papers where we propose a method to find such sources. The structure of this article is organized as follows: In Sec. \ref{S:Reg_Sol}, we present the asymptotically locally flat solution in 2+1 dimensions and discuss the properties of the regularized spacetime. The field sources are studied in Sec. \ref{S:Field_Source}. In Sec. \ref{S:EC}, we examine which of the energy conditions are violated. In Sec. \ref{S:Geo}, we analyze the trajectories of both massive and massless particles in the regularized spacetime. Our conclusions and perspectives are presented in Sec. \ref{S:conclusions}.

\section{The regularized solution}\label{S:Reg_Sol}
Just as has been done with several solutions in the literature \cite{Franzin:2021vnj}, we can apply the Simpson-Visser regularization method to eliminate the presence of singularity. The regularized solution is
\begin{equation}
    ds^2=f(r)dt^2-f(r)^{-1}dr^2-\Sigma^2\left(r\right)d\varphi^2,\label{line_reg}
\end{equation}
with
\begin{equation}
    f(r)=-\Lambda (r^2+a^2)+b\sqrt{r^2+a^2}-\mu, \quad \mbox{and} \quad \Sigma(r)=\sqrt{r^2+a^2}.
\end{equation}
From now on, we will analyze the locally flat and AdS case separately.

\subsection{Asymptotically locally flat black hole}
For $\Lambda=0$, the regularized solution is
\begin{equation}
    f(r)=b\sqrt{r^2+a^2}-\mu, \quad \mbox{and} \quad \Sigma(r)=\sqrt{r^2+a^2}.
\end{equation}
The curvature invariants to the regularized spacetime are
\begin{eqnarray}
    R&=&\frac{b \left(3 a^2+2 r^2\right)}{\left(a^2+r^2\right)^{3/2}}-\frac{2 a^2 \mu }{\left(a^2+r^2\right)^2},\\
    \mathcal{K}&=&-\frac{4a^2 \mu b\left(2 a^2 + r^2\right)}{\left(a^2+r^2\right)^{7/2}}+\frac{4 a^4 \mu ^2}{\left(a^2+r^2\right)^4}+\frac{b^2  \left(5 a^4+4 a^2 r^2+2 r^4\right)}{\left(a^2+r^2\right)^3}.
\end{eqnarray}
The invariants become more complex than in the singular case, but if we analyze the asymptotic behaviors, we observe that to $r\rightarrow 0$ we find that both $R$ and $\mathcal{K}$ tend to a constant. To $r\rightarrow \infty$, we find that $R\propto r^{-1}$ and $\mathcal{K}\propto r^{-2}$. It means that the spacetime has no curvature singularities and is asymptotically locally flat.

Solving $f(r)=0$, we find that the radius of the event horizon is given by $r_h=\sqrt{\mu^2/b^2-a^2}$. Depending on the values chosen for the parameter $a$, the properties of the spacetime structure can be altered. In summary, we have that:
\begin{itemize}
    \item For $a < \mu/b$, we have a RBH with two horizons. At $r=0$, the circumference of this solution, $\mathcal{C}=2\pi \sqrt{r^2+a^2}$, has a minimum, which is precisely the BB.

    \item 
If $a=\mu/b$, the event horizon and the minimum circumference are located at the same point, and in this way, we have a black throat.

\item 
If $a>\mu/b$, there are no longer any horizons, and thus we have a TWH in $2+1$ dimensions with the throat located at $r=0$.

\item If $\mu=0$, the event horizon radius would have an imaginary value. In this case, there are no horizons present, and thus we do not have a black hole but rather a WH.

\end{itemize}

\subsection{Asymptotically AdS black hole}
In the case where we have $\Lambda=-1/l^2$, the regularized solution is given by
\begin{equation}
    f(r)=\frac{1}{l^2}(r^2+a^2)+b\sqrt{r^2+a^2}-\mu, \quad \mbox{and} \quad \Sigma(r)=\sqrt{r^2+a^2}.
\end{equation}
 The curvature invariants to the regularized spacetime are
\begin{eqnarray}
    R&=&\frac{4 a^2+6 r^2}{l^2 \left(a^2+r^2\right)}+\frac{b \left(3 a^2+2 r^2\right)}{\left(a^2+r^2\right)^{3/2}}-\frac{2 a^2 \mu }{\left(a^2+r^2\right)^2},\\
    \mathcal{K}&=&\frac{4 \left(2 a^4+4 a^2 r^2+3 r^4\right)}{l^4 \left(a^2+r^2\right)^2}-\frac{8 a^2 \mu }{l^2 \left(a^2+r^2\right)^2}+\frac{4 b \left(3 a^4+4 a^2 r^2+2 r^4\right)}{l^2 \left(a^2+r^2\right)^{5/2}}\nonumber\\
    &-&\frac{4a^2 \mu b\left(2 a^2 + r^2\right)}{\left(a^2+r^2\right)^{7/2}}+\frac{4 a^4 \mu ^2}{\left(a^2+r^2\right)^4}+\frac{b^2  \left(5 a^4+4 a^2 r^2+2 r^4\right)}{\left(a^2+r^2\right)^3}.
\end{eqnarray}
The invariants become more complex than in the singular case, but if we analyze the asymptotic behaviors, we observe that for $r\rightarrow 0$ we find that both $R$ and $\mathcal{K}$ tend to a constant. To $r\rightarrow \infty$, we find that $R\propto \frac{6}{l^2}+ 2b r^{-1}$ and $\mathcal{K}\propto \frac{12}{l^4}+\frac{8b}{l^2 r}$. It means that the spacetime has no curvature singularities and is asymptotically AdS.

Solving $f(r)=0$, we find four possibilities that are given by
\begin{equation}
   r=\pm \sqrt{\left(\frac{bl^2}{2}\pm\sqrt{\frac{b^2l^4}{4}+l^2\mu}\right)^2-a^2}.
\end{equation} 
As said above, if $b$ and $\mu$ are positive, the structure is identical to that of charged BB spacetime studied in Ref. \cite{Franzin:2021vnj}. However, the fact that the parameters can be negative will give rise to new possibilities, described below:
\begin{enumerate}
   \item $b > 0$ and $\mu\leq 0$: There is no event horizon and we have a TWH with a throat at $r=0$.
    \item $b > 0$ and $\mu>0$: This analysis is identical to the charged BB, but we repeat here for completes. 
  \begin{itemize}
     \item For $a>bl^2/2 \pm \sqrt{b^2l^4/4+l^2\mu}$, we have no horizon. However, since there is no singularity, we have a TWH. 
      \item For $a=bl^2/2 \pm \sqrt{b^2l^4/4+l^2\mu}$, the event horizon and the minimum circumference are located at the same point, and in this way, we have a black throat.
      \item For $a < bl^2/2 \pm \sqrt{b^2l^4/4+l^2\mu}$, we have a wormhole (WH) throat inside an event horizon.
  \end{itemize} 
 \item $b < 0$ and $\mu=0$: We have a event horizon located at $r=\pm \sqrt{b^2l^4-a^2}$.
  \item $b < 0$ and $\mu>0$: Depending on the value of $a$, we have one or no event horizon.
  \item $b < 0$ and $ -b^2l^2/4<\mu<0$: The WH throat is surrounded by a Cauchy horizon and we also have an event horizon.
   \item $b < 0$ and $ -b^2l^2/4=\mu$: The WH throat is surrounded by an extremal horizon.
   \item  $b < 0$ and $\mu< -b^2l^2/4$: There is no horizon, and we have a TWH with a throat at $r=0$.
\end{enumerate}

\section{Field sources}\label{S:Field_Source}
In general, regular solutions can arise in various modified theories of gravity. However, the asymptotically locally flat black hole solution originally emerged as a solution to the field equations of $K$--gravity, which is a specific case of modified gravity theories of the type $f(R,R_{\mu\nu})$. The regular spacetime is proposed in a generic manner without considering the gravitational theory that generates it. However, it makes more sense to work within the same theory to study the properties of the regularized solution.
So, let us consider that this metric is a solution of the $K$--gravity in $2+1$ dimensions and determine the material content that can generate it. To do this, we will consider the theory described by the action:
\begin{equation}
    S=\int \sqrt{\left|g\right|}d^3x\left[K - 2 h\left(\phi\right) g^{\mu\nu}\partial_\mu \phi\partial_\nu \phi
   		+2V(\phi) + L(F)\right],    			\label{Action}
\end{equation}
where $\phi$ is the scalar field, $V(\phi)$ is its potential, $L(F)$ is the NED Lagrangian, $F=F^{\mu\nu}F_{\mu\nu}$ is the electromagnetic scalar, and $F_{\mu\nu} = \partial_\mu A_\nu - \partial_\nu A_\mu$ is the electromagnetic field tensor. The function $h\left(\phi\right)$ will determine if the scalar field is phantom, $h\left(\phi\right)<0$, or standard, $h\left(\phi\right)>0$.
  
Varying the action \eqref{Action} with respect to $\phi$, $A_\mu$, and $g^{\mu\nu}$, we obtain the field equations
\begin{eqnarray}
    &&\nabla_\mu \left[L_F F^{\mu\nu}\right] =\frac{1}{\sqrt{\left|g\right|}}\partial_\mu \left[\sqrt{\left|g\right|}L_F F^{\mu\nu}\right]=0,\label{eq-F}\\
     &&2h\left(\phi\right) \nabla_\mu \nabla^\mu\phi +\frac{dh\left(\phi\right)}{d\phi}\partial^\mu\phi\partial_\mu\phi=-  \frac{dV(\phi)}{d\phi} ,     \label{eq-phi}\\  
      &&K_{\mu\nu} =2\square R_{\mu\nu}-\frac{1}{2}\nabla_\mu\nabla_\nu R-\frac{1}{2}g_{\mu\nu}\square R+4R_{\mu\rho\nu\sigma}R^{\rho\sigma}-g_{\mu\nu}R_{\rho\sigma}R^{\rho\sigma}
-\frac{3}{2}RR_{\mu\nu}+\frac{3}{8}g_{\mu\nu}R^2\nonumber\\
&& =2T[\phi]_{\mu\nu}  +2 T[F]_{\mu\nu},				\label{eq-Ein}
\end{eqnarray}
where $L_F=d L/d F$, $T[\phi]_{\mu\nu}$ and $T[F]_{\mu\nu}$ are the stress-energy tensors of the scalar and electromagnetic fields, respectively,
\begin{eqnarray}                
   T[F]_{\mu\nu} = \frac{1}{2} g_{\mu\nu} L(F ) - 2L_F {F_{\nu}}^{\alpha} F_{\mu \alpha},\label{SETF}\\
    T[\phi]_{\mu\nu} =  2 h\left(\phi\right) \partial_\nu\phi\partial_\mu\phi 
   - g_{\mu\nu} \big (h\left(\phi\right) \partial^\alpha \phi \partial_\alpha \phi - V(\phi)\big).\label{SETphi}
\end{eqnarray}  

It is important to emphasize that the stress-energy tensors for the scalar field and for NED have the symmetries $T[\phi]^0_0 = T[\phi]^2_2$ and $T[F]^0_0 = T[F]^1_1$, respectively. However, from equation \eqref{eq-Ein}, it is possible to show that, for a line element of the form \eqref{line_reg}, the tensor $K^\mu_\nu$ does not have these symmetries. Therefore, a scalar field alone or a NED alone cannot serve as the source for this solution. This justifies the need for the choice of sources and their combination.

Solving the Maxwell equations, considering an electrically charged solution, we find that the electric field has the form
\begin{equation}
    F^{10}=\frac{qL_F^{-1}}{\sqrt{r^2+a^2}},
\end{equation}
and the scalar $F$ is
\begin{equation}
    F=-\frac{2q^2}{ L_F^2 (r^2+a^2)}.
\end{equation}

Using the $1-1$ and $2-2$ components of the gravitational field equations, we can obtain that the electromagnetic quantities are given by
\begin{eqnarray}            
 		L(r)&=&\frac{3}{4} f f^{(4)}-\frac{1}{8}f''^2+\frac{1}{2}\frac{ f \left(8 f' \Sigma ' \Sigma ''+f \left(5 \Sigma
   ''^2+4 \Sigma ^{(3)} \Sigma '\right)\right)}{\Sigma ^2}+\frac{1}{2} f^{(3)} f'\nonumber\\
   &-&\frac{ \left(22 f \Sigma
   ^{(3)} f'+8 f'^2 \Sigma ''+f \left(-2 f^{(3)} \Sigma '+13 f'' \Sigma ''+6 f \Sigma
   ^{(4)}\right)\right)}{4\Sigma }-\frac{2 f^2 \Sigma '^2 \Sigma ''}{\Sigma ^3}-2 V,\label{T2D02}
   \\
  L_F(r)&=&16 q^2 \Sigma \left\{\Sigma ^2 \left(\Sigma  f''^2+12 f'^2 \Sigma ''-f'
   \left(3 f^{(3)} \Sigma +f'' \Sigma '\right)\right)\right.\nonumber\\
   &+&2 f \Sigma  \left(\Sigma
    \left(-f^{(4)} \Sigma +f^{(3)} \Sigma '+3 f'' \Sigma ''+9 \Sigma
   ^{(3)} f'+8 \Sigma h  \phi '^2\right)-12 f' \Sigma '\Sigma
   ''\right)\nonumber\\
   &+&\left.4 f^2 \left(\Sigma  \left(\Sigma  \Sigma ^{(4)}-\Sigma
   ''^2\right)-3 \Sigma  \Sigma ^{(3)} \Sigma '+3 \Sigma '^2 \Sigma ''\right)\right\}^{-1}.\label{T2D01+}
\end{eqnarray}  
To obtain the electromagnetic expressions, we must first derive the quantities associated with the scalar field. Solving the component $0-0$ of the gravitational field equations, it's possible to find $h(\phi) \phi'^2$ in terms of the metric functions. Following the procedure developed by \cite{Bronnikov:2022bud}, we will choose the scalar field as a monotonic function
\begin{equation}
    \phi=\arctan \left(\frac{r}{a}\right), \qquad \phi \in \left(-\pi/2,\pi/2\right),
\end{equation}
and then we solve the component $0-0$ of the gravitational field equations to find $h\left(\phi\right)$, that is given by
\begin{equation}
    h\left(\phi\right)=\frac{a^2 \left(5 \mu -b \sqrt{a^2+r^2}\right)-12 \mu  r^2}{4 \left(a^2+r^2\right)^2}.
\end{equation}

The choice of the scalar field may seem somewhat arbitrary. However, this scalar field is regular at $r=0$ and bounded, not growing indefinitely. This scalar field model also arises naturally in general relativity when studying some WH and BB models \cite{Bronnikov:1973fh,Alencar:2024yvh}, in addition to allowing an analytical expression for $h(\phi)$ to be obtained. There are also other scalar field models that exhibit some of these characteristics. In \cite{Crispim:2024lzf}, the authors consider several models for $\phi$ to study WH solutions, but conclude that this choice does not affect the electromagnetic functions and does not drastically change the regions where the field is phantom or canonical. In \cite{Pereira:2024rtv}, the authors consider a $k$-essence scalar field for the Simpson-Visser case and verify that this change in the type of scalar field also does not alter the form of the expressions related to NED. We also find these same properties here.

\begin{figure}
    \centering
    \includegraphics[width=0.6\linewidth]{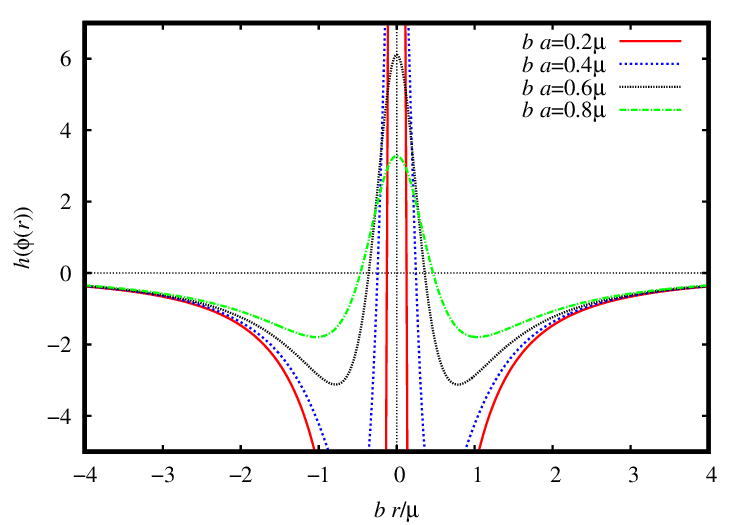}
    \caption{Behavior of the function $h(\phi)$
as a function of the radial coordinate for different values of the regularization parameter.}
    \label{fig:hphi}
\end{figure}

In Fig. \ref{fig:hphi}, we observe the behavior of the function $h(\phi)$ and notice that it changes sign depending on the region. As a result, the scalar field is canonical in some regions and phantom in others, making it a partly phantom scalar field.

We still need to determine $V(r)$. To do this, we must solve the equation \eqref{eq-phi}, which results in
\begin{equation}
    V(r)=\frac{a^2 b \mu  \left(197 a^2-168 r^2\right)}{140 \left(a^2+r^2\right)^{7/2}}-\frac{a^2 \left(2
   a^4 b^2+a^2 \left(2 b^2 r^2+19 \mu ^2\right)-32 \mu ^2 r^2\right)}{16 \left(a^2+r^2\right)^4}.
\end{equation}
Once we have the expressions related to the scalar field, we can obtain the electromagnetic quantities, which are
\begin{eqnarray}
   L(r)=\frac{132 a^2 b \mu  \left(7 r^2-3 a^2\right)}{35 \left(a^2+r^2\right)^{7/2}}+\frac{a^2 \left(15
   \mu ^2 \left(5 a^2-16 r^2\right)+b^2 \left(13 a^4+a^2 r^2-12 r^4\right)\right)}{8
   \left(a^2+r^2\right)^4},\\
    L_F(r)=-\frac{16 q^2 \left(a^2+r^2\right)^{3}}{a^2  \left(b^2 \left(7 a^2-8 r^2\right)
   \left(a^2+r^2\right)+36 \mu ^2 \left(a^2-4 r^2\right)\right)-44 a^2 b \mu  \left(a^2-3
   r^2\right) \sqrt{a^2+r^2}}.
\end{eqnarray}
Despite being obtained as independent functions, the electromagnetic quantities are related to each other and must satisfy the consistency relation
  \begin{equation}
     L_F-\frac{dL}{dr}\left(\frac{dF}{dr}\right)^{-1}=0,\label{Cons}
  \end{equation}
which is indeed satisfied.

Since we are dealing with electrically charged sources, it is not possible to write the expression for $L(F)$ analytically. However, we can still explicitly express $V(\phi)$ and $h(\phi)$, which are given by:
\begin{eqnarray}
    V(\phi)&=&\frac{a^4 b \mu  \left(197-168 \tan ^2\phi\right)}{280 \left(a^2 \sec ^2\phi
   \right)^{7/2}}-\frac{\cos ^6\phi \left(4 a^2 b^2+51 \mu ^2 \cos 2
   \phi-13 \mu ^2\right)}{32 a^4},\\
   h(\phi)&=&-\frac{\cos ^4(\phi) \left(b \sqrt{a^2 \sec ^2(\phi)}+12 \mu  \tan ^2(\phi)-5 \mu
   \right)}{4 a^2}.
\end{eqnarray}

In this way, we were able to find the quantities associated with the matter fields of the source such that all the field equations are satisfied.

\section{Energy conditions}\label{S:EC}
In this section, we will analyze the energy conditions of the regularized asymptotically locally flat solution. The field equations of the purely quadratic gravity can be rewritten as
\begin{equation}
\label{EoM}
2\square R_{\mu\nu}-\frac{1}{2}\nabla_\mu\nabla_\nu R-\frac{1}{2}g_{\mu\nu}\square R+4R_{\mu\rho\nu\sigma}R^{\rho\sigma}-g_{\mu\nu}R_{\rho\sigma}R^{\rho\sigma}
-\frac{3}{2}RR_{\mu\nu}+\frac{3}{8}g_{\mu\nu}R^2=2T_{\mu\nu},
\end{equation}
where $T_{\mu\nu}$ is the stress-energy tensor. 

In order to analyze the energy conditions associated with the regularized solution, we rewrite the stress-energy tensor considering that the matter is described by an anisotropic fluid. In regions where $f(r)>0$, we can write the stress-energy tensor as
\begin{equation}
    {T^{\mu}}_{\nu}=\mbox{diag}\left[\rho,\, -p_r,\, -p_t\right],
\end{equation}
where $\rho$ is the energy density, $p_r$ is the radial pressure, and $p_t$ is the tangential pressure. In regions where $f(r)<0$, the stress-energy tensor is written as
\begin{equation}
    {T^{\mu}}_{\nu}=\mbox{diag}\left[-p_r,\, \rho,\, -p_t\right].
\end{equation}

From the field equations \eqref{EoM}, we can immediately obtain the expressions for the fluid quantities. We have for $f(r)>0$:
\begin{equation}
\rho = \frac{a^2 \left(4 \mu ^2 \left(32 r^2-9 a^2\right)-b^2 \left(7 a^2-4 r^2\right) \left(a^2+r^2\right)\right)}{16
   \left(a^2+r^2\right)^4}+\frac{a^2 b \mu  \left(11 a^2-30 r^2\right)}{4 \left(a^2+r^2\right)^{7/2}},
\end{equation}

\begin{equation}
-p_r  = \frac{a^2 \left(b^2 \left(a^2+r^2\right) \left(a^2+4 r^2\right)+4 \mu ^2 \left(a^2+8 r^2\right)\right)}{16
   \left(a^2+r^2\right)^4}-\frac{a^2 b \mu  \left(a^2+6 r^2\right)}{4 \left(a^2+r^2\right)^{7/2}},
\end{equation}

\begin{eqnarray}
-p_{t} &=& \frac{a^2 \left(b^2 \left(7 a^2-12 r^2\right) \left(a^2+r^2\right)+4 \mu ^2 \left(9 a^2-40 r^2\right)\right)}{16
   \left(a^2+r^2\right)^4}+\frac{a^2 b \mu  \left(36 r^2-11 a^2\right)}{4 \left(a^2+r^2\right)^{7/2}}.
\end{eqnarray}
The above expressions for energy density and radial pressure must be exchanged if $f(r)<0$. The analysis will become clearer if we introduce a new radial coordinate $\tilde{r}^2=r^2+a^2$. The fluid quantities written in terms of $\tilde{r}$ are given by:
\begin{equation}
\label{rhotilde}
\rho(\tilde{r}) =\frac{a^2b^2}{16 \tilde{r}^8} \left[
4 \tilde{r}^4
- 120 \frac{\mu}{b} \tilde{r}^3
+ \left(-11 a^2 + 128 \frac{\mu^2}{b^2} \right) \tilde{r}^2
+ 164 a^2 \frac{\mu}{b} \tilde{r}
- 164 a^2 \frac{\mu^2}{b^2}
\right],
\end{equation}

\begin{equation}
\label{prtilde}
-p_r(\tilde{r}) = \frac{a^2 b^2}{16 \tilde{r}^8} \left[
4 \tilde{r}^4
- 24 \frac{\mu}{b} \tilde{r}^3
+ \left(-3 a^2 + 32 \frac{\mu^2}{b^2} \right) \tilde{r}^2
+ 20 a^2 \frac{\mu}{b} \tilde{r}
- 28 a^2 \frac{\mu^2}{b^2}
\right],
\end{equation}

\begin{equation}
\label{pttilde}
-p_t(\tilde{r}) = \frac{a^2 b^2}{16\tilde{r}^8} \left[
-12 \tilde{r}^4
+ 144 \frac{\mu}{b} \tilde{r}^3
+ \left(19 a^2 - 160 \frac{\mu^2}{b^2} \right) \tilde{r}^2
- 188 a^2 \frac{\mu}{b} \tilde{r}
+ 196 a^2 \frac{\mu^2}{b^2}
\right].
\end{equation}

Once we have identified the fluid quantities, the energy conditions associated with the stress-energy tensor are given by
\begin{eqnarray}
&&NEC_{1,2}=WEC_{1,2}=SEC_{1,2} 
\Longleftrightarrow \rho+p_{r,t}\geq 0,\label{Econd1} \\
&&SEC_3 \Longleftrightarrow\rho+p_r+p_t\geq 0,\label{Econd2}\\
&&DEC_{1,2} \Longleftrightarrow \rho-|p_{r,t}|\geq 0 \Longleftrightarrow 
(\rho+p_{r,t}\geq 0) \hbox{and} (\rho-p_{r,t}\geq 0),\label{Econd3}\\
&&DEC_3=WEC_3 \Longleftrightarrow\rho\geq 0.\label{Econd4}
\end{eqnarray}
As part of the dominant energy condition is already in the null energy condition, we will consider only $DEC_{1,2}\Longrightarrow \rho-p_{r,t}\geq 0$. In the notation we are following, the conditions $NEC$, $WEC$, $DEC$, and $SEC$ are the null, weak, dominant, and strong energy conditions, respectively. Subscript 1 is used when the energy density with radial pressure is considered, and subscript 2 when the energy density with tangential pressure is used. Subscript 3 is used when only the energy density is considered, as in the case of $WEC_3$, or the energy density with both pressures, as in the case of $SEC_3$. Finally, we see that all the conditions will depend on the relationship between these parameters $a$, $b$, and $\mu$. Therefore, we will consider the cases $a<\mu/b$, where the metric \eqref{line_reg} will describe a RBH with two horizons, $a = \mu/b$, describing a RBH with a black throat, and finally the case $a>\mu/b$, describing a WH with a throat located in $r=0$ ($\tilde{r} = a$).

From equations \eqref{rhotilde}, \eqref{prtilde}, and \eqref{pttilde}, we can analyze the validity of each of the energy conditions mentioned above. The following restriction gives $WEC_3$:
\begin{equation}
\rho(\tilde{r}) =\frac{a^2b^2}{16 \tilde{r}^8} N_1(\tilde{r}) \geq 0\;; N_1(\tilde{r})=
4 \tilde{r}^4
- 120 \frac{\mu}{b} \tilde{r}^3
+ \left(-11 a^2 + 128 \frac{\mu^2}{b^2} \right) \tilde{r}^2
+ 164 a^2 \frac{\mu}{b} \tilde{r}
- 164 a^2 \frac{\mu^2}{b^2}.
\end{equation}
As we can see, the energy density is an eighth-degree polynomial in the denominator and a fourth-degree polynomial in the numerator. Since the denominator is always positive, it will be $N_1(\tilde{r})$ that determines the sign of $\rho(\tilde{r})$.  

\begin{figure}
    \centering
    \includegraphics[scale=0.65]{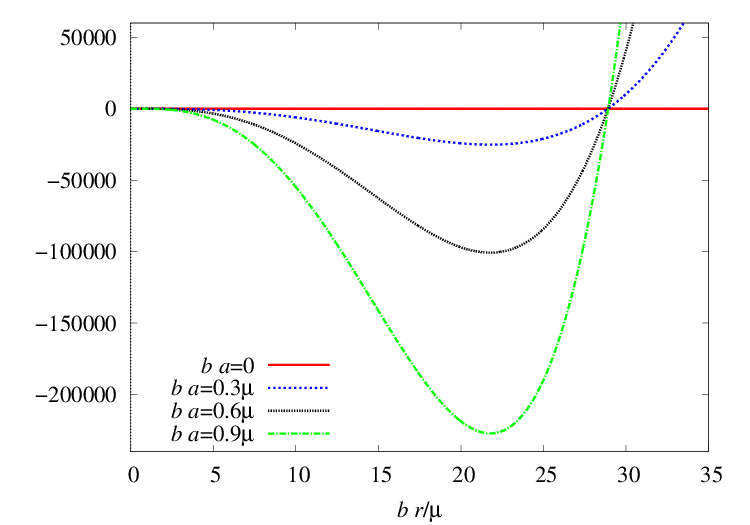}
    \includegraphics[scale=0.65]{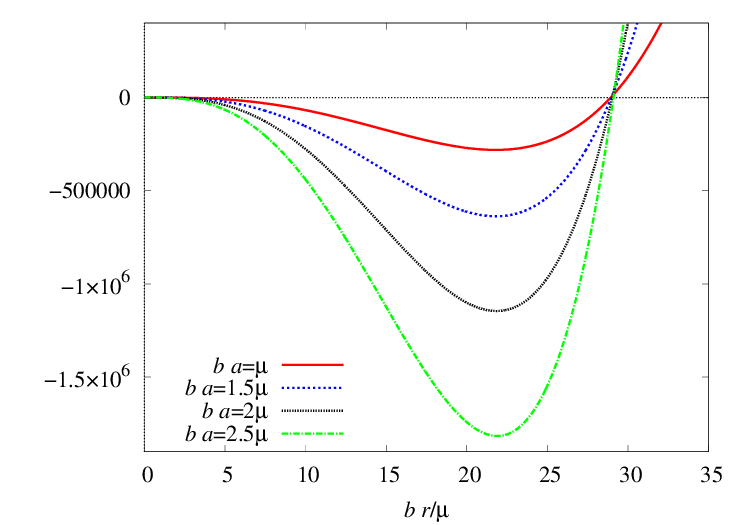}
    \caption{The numerator of the energy density considering different values of the parameter $a$.}
    \label{Fig-EnergyDensity}
\end{figure}

Let us now study whether, for some range of parameters, $WEC_3$ can be satisfied. First, for regions far away, both in the case of RBH and WH, the energy density will be positive. On the other hand, at $r=0$, we have
\begin{equation}
   N_1(a)=- 7a^2\left(a^2-\frac{44 a \mu }{7 b}+\frac{36 \mu ^2}{7 b^2}\right).
\end{equation}
We see, therefore, that the energy density is positive for 
\begin{equation}\label{rangeN1}
    \frac{2  (11-\sqrt{58})}{7 }\frac{\mu}{b} \leq a\leq\frac{2  (11+\sqrt{58})}{7 }\frac{\mu}{b}.
\end{equation}
The second derivative of $N_1$ is given by 
\begin{equation}
N''_1(\tilde{r})=   48 \, \tilde{r}^2 - \frac{720 \, \tilde{r} \, \mu}{b} + 2 \left( -11 a^2 + \frac{128 \mu^2}{b^2} \right).
\end{equation}
The interesting thing about the above expression is that its roots are given by 
\begin{equation}
\tilde{r}_{\pm} = \frac{1}{12} \left( \frac{90 \mu}{b} \pm \frac{\sqrt{6} \sqrt{11 a^2 b^2 + 1222 \mu^2}}{b} \right).
\end{equation}
From the above roots, we can directly find $N_1(\tilde{r}_+)<0$. Now, remember that the limit of \( N_1(\tilde{r}_1) \) as \( \tilde{r}_1 \to \infty \) is \( +\infty \). Therefore, a necessary condition for $WEC_3$ to be satisfied is that the throat, $\tilde{r}=a$, is after this point. That is, we must have 
\begin{equation}
\frac{1}{12} \left( \frac{90 \mu}{b} + \frac{\sqrt{6} \sqrt{11 a^2 b^2 + 1222 \mu^2}}{b} \right) < a.
\end{equation}
The above condition can be solved, and we get
\begin{equation}
a < \frac{14 \mu}{39 b} \quad \text{or} \quad a > \frac{1066 \mu}{39 b}.
\end{equation}
However, the above condition is outside the range (\ref{rangeN1}), which guarantees that $WEC_3$ is satisfied at the throat. We conclude that for any value of the parameters, there will be some region in which $WEC_3$ is violated.  To visualize this, the behavior of $ N_1(\tilde{r})$, for different values of the parameters, can be seen in Fig. \ref{Fig-EnergyDensity}.   A source of energy density originates from a type of matter known as exotic matter. Exotic matter is particularly important in the context of WHs, as according to the GR, Morris-Thorne WHs would require this type of matter to exist. We can observe that exotic matter capable of supporting such spacetimes can also appear in $K$-gravity.

\begin{figure}
    \centering
    \includegraphics[scale=0.65]{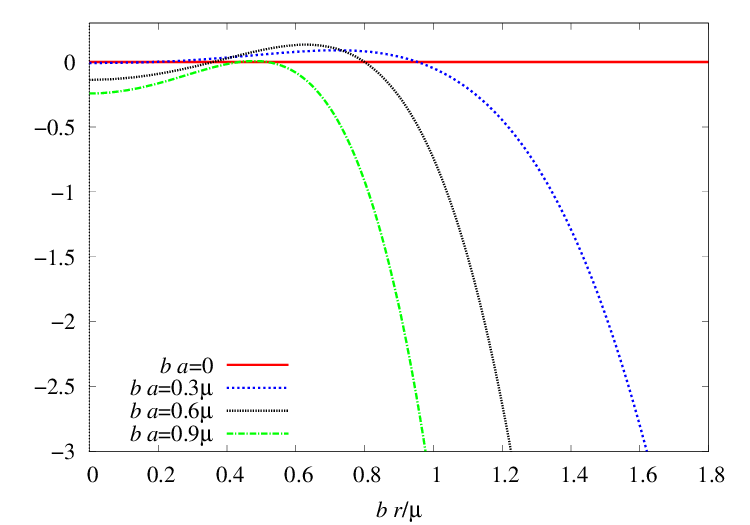}
    \includegraphics[scale=0.65]{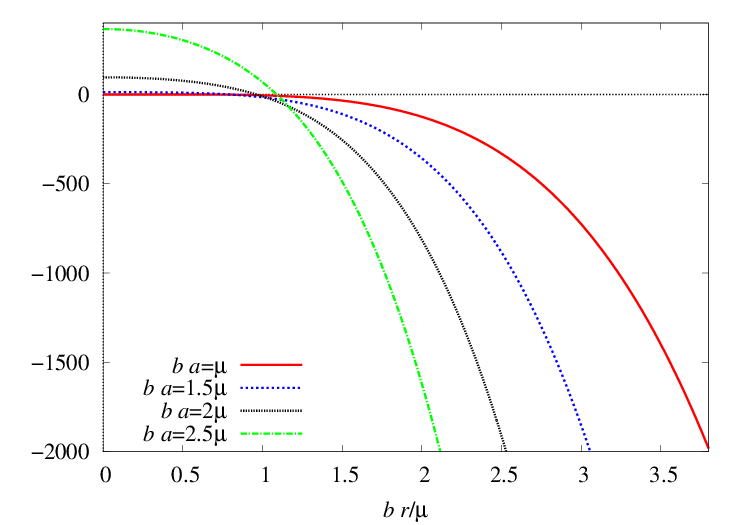}
    \caption{Validity of the $WEC_1$ considering different values of the parameter $a$.}
    \label{Fig-WEC1}
\end{figure}

\begin{figure}
    \centering
    \includegraphics[scale=0.65]{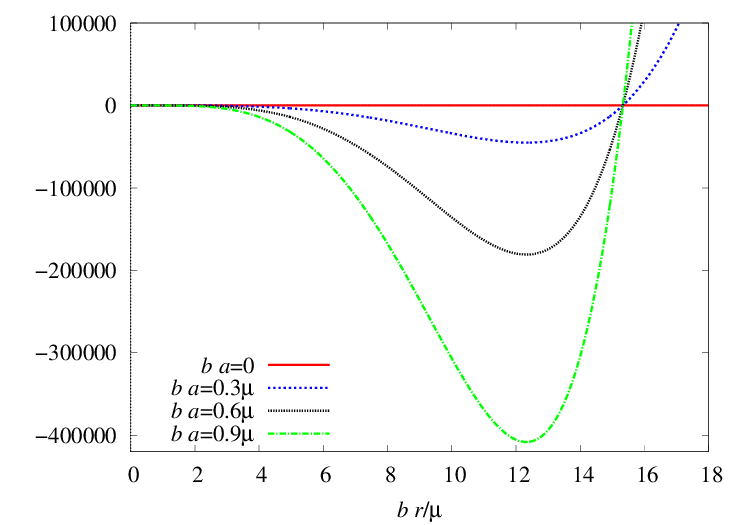}
    \includegraphics[scale=0.65]{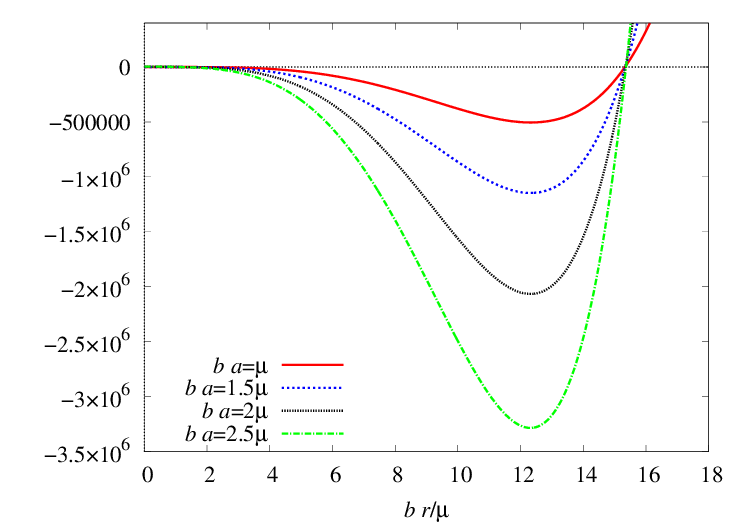}
    \caption{Validity of the $WEC_2$ considering different values of the parameter $a$.}
    \label{Fig-WEC2}
\end{figure}

Next, we will check the validity of $NEC_{1,2}=WEC_{1,2}=SEC_{1,2}$. We must have for $WEC_1$:
\begin{equation}
\rho + p_r = -\frac{a^2 (b \tilde{r}-\mu ) \left(a^2 (b \tilde{r}-17 \mu )+12 \mu  \tilde{r}^2\right)}{2 \tilde{r}^8}\geq 0.
\end{equation}
The above expression is always negative at large $\tilde{r}$, for any parameters. Therefore, $WEC_1$ will be violated in some regions, as can also be seen in Fig. \ref{Fig-WEC1}.
For $WEC_2$, we have:
\begin{equation}
\rho + p_t = \frac{a^2b^2}{16 \tilde{r}^8}N_2(\tilde{r})\geq 0\;; N_2(\tilde{r})=16 \tilde{r}^4 
- \frac{264 \mu}{b} \tilde{r}^3 
+ \left( -30 a^2 + \frac{288 \mu^2}{b^2} \right) \tilde{r}^2 
+ \frac{352 a^2 \mu}{b} \tilde{r} 
- \frac{360 a^2 \mu^2}{b^2}.
\end{equation}   
Therefore, we get that, for large $\tilde{r}$, $WEC_2$ is always satisfied. At $\tilde{r}=a$, we have
\begin{equation}
  N_2(a) =-14a^2\left(a^2-\frac{44 a \mu }{7 b}+\frac{36 \mu ^2}{7 b^2}\right),
\end{equation}
with roots
\[
a = \frac{2 \mu}{7 b} \left( 11 \pm \sqrt{58} \right).
\]
Interestingly, the range of parameters in which $WEC_2$ is valid at the throat is identical to the one found for $WEC_3$. Although very similar to the case of $WEC_3$, we were unable to obtain analytical conclusions about the violation of $WEC_2$.  For example, in this case, we have  
\begin{equation}
N''_2(\tilde{r})=192 \tilde{r}^2 
- \frac{1584 \tilde{r} \mu}{b} 
+ \frac{576 \mu^2}{b^2} 
- 60 a^2.
\end{equation}
The roots are
\begin{equation}
    \tilde{r}_{\pm} = \frac{1}{8} \left( \frac{33 \mu}{b} \pm \frac{\sqrt{20 a^2 b^2 + 897 \mu^2}}{b} \right),
\end{equation}
and we obtain $N_2(\tilde{r}_+)>0$. Therefore, we cannot impose a second condition. However, $N_2(\tilde{r})$ is plotted in Fig. \ref{Fig-WEC2}, for the cases where $WEC_2$ is satisfied at the throat, and we can see that, even so, some regions will violate $WEC_2$.  

\begin{figure}
    \centering
    \includegraphics[scale=0.65]{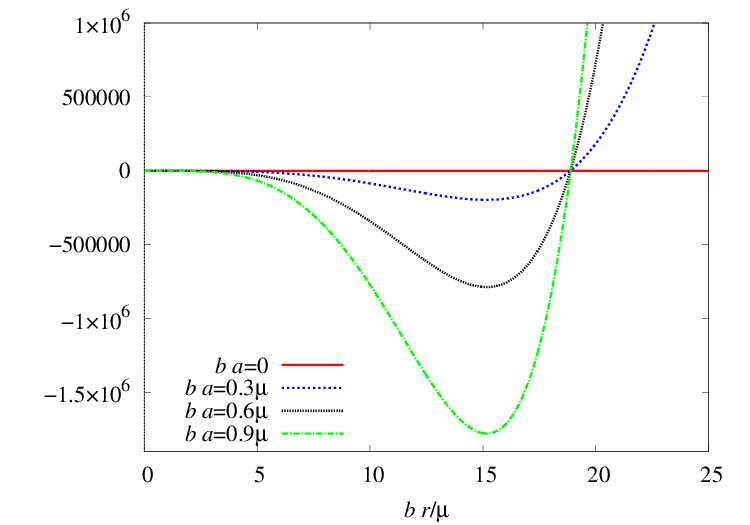}
    \includegraphics[scale=0.65]{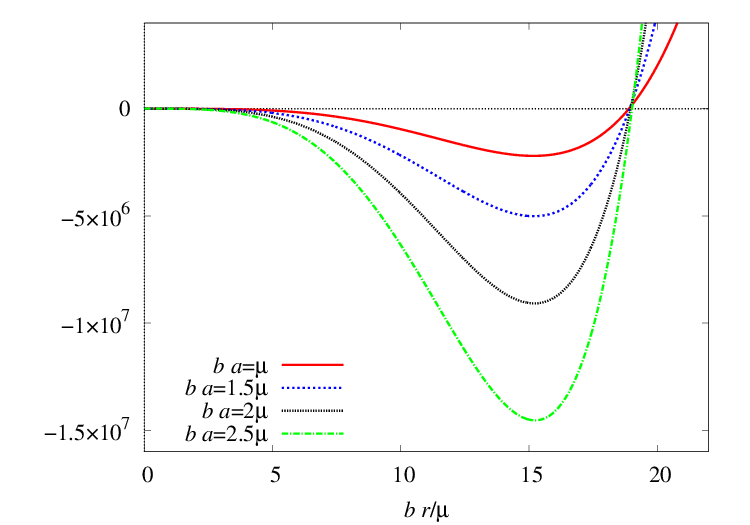}
    \caption{Validity of the $SEC_3$ considering different values of the parameter $a$.}
    \label{Fig-SEC3}
\end{figure}

Now, we will check if $SEC_3$ is satisfied. The strong energy condition is satisfied if we have
\begin{align}
\rho + p_r + p_t &= \frac{a^2 b^2}{16 \tilde{r}^8} N_3(\tilde{r}) \geq 0 \label{eq:nec}; \\
N_3(\tilde{r}) &= 12 \tilde{r}^4 
- \frac{240 \mu}{b} \tilde{r}^3 
+ \left( -27 a^2 + \frac{256 \mu^2}{b^2} \right) \tilde{r}^2 
+ \frac{332 a^2 \mu}{b} \tilde{r} 
- \frac{332 a^2 \mu^2}{b^2} \nonumber.
\end{align}
We see that for large $\tilde{r}$ it is satisfied and 
\begin{equation}
 N_3(a) =  -15 a^2\left(a^2-\frac{92 a \mu }{15 b}+\frac{76 \mu ^2}{15 b^2}\right).
\end{equation}
Therefore, $SEC_3$ is satisfied in the range  
\begin{equation}
\frac{2 \mu}{15 b} \left( 23 - 2 \sqrt{61} \right) 
< a < 
\frac{2 \mu}{15 b} \left( 23 + 2 \sqrt{61} \right).
\end{equation}
The second derivative of $N_3$ is given by 
\begin{equation}
N''_3(\tilde{r})= 144 \tilde{r}^2 
- \frac{1440 \mu}{b} \tilde{r} 
+ 2 \left( \frac{256 \mu^2}{b^2} - 27 a^2 \right).
\end{equation}
The roots are given by 
\begin{equation}
\tilde{r}_{\pm} = \frac{1}{12 b} \left[ 60 \mu \pm \sqrt{2} \sqrt{27 a^2 b^2 + 1544 \mu^2} \right],
\end{equation}
and we can find directly that $ N_1(\tilde{r}_+)<0$. By following the same reasoning as before, we get the condition
\begin{equation}
a < \frac{\mu}{90 b} \left( 720 - \sqrt{472320} \right)
\quad \text{or} \quad
a > \frac{\mu}{90 b} \left( 720 + \sqrt{472320} \right).
\end{equation}
Since both conditions cannot be satisfied for any range of parameters, we find that there will always be some region in which $SEC_3$ will be violated. This can also be seen in Fig. \ref{Fig-SEC3}, for different values of the parameter $a$.

Finally, we will consider the dominant energy conditions $DEC_1$ and $DEC_2$. We can immediately notice that $DEC_{1,2}$ encompasses $WEC_{1,2}$, once for the inequality $\rho-|p_{r,t}|\geq 0$ to be true, we must have $\rho+p_{r,t}\geq0$ which is precisely $WEC_{1,2}$. However, we still need to verify the second inequality for each of these solutions to determine whether the dominant energy condition is indeed satisfied.

\begin{figure}
    \centering
    \includegraphics[scale=0.65]{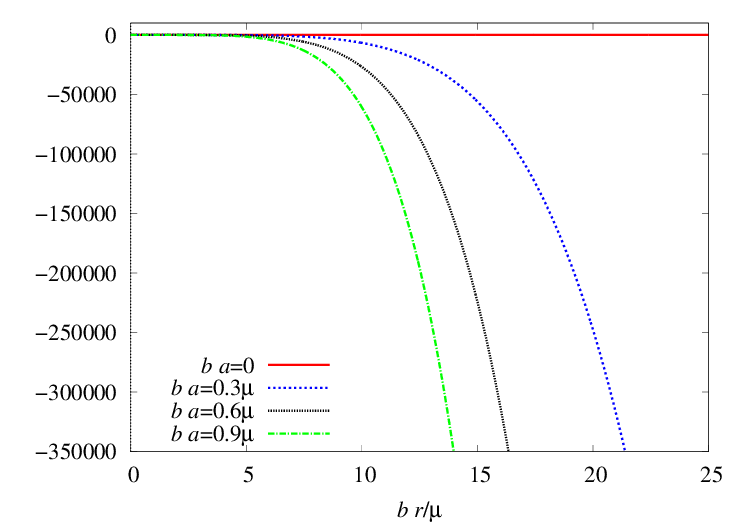}
    \includegraphics[scale=0.65]{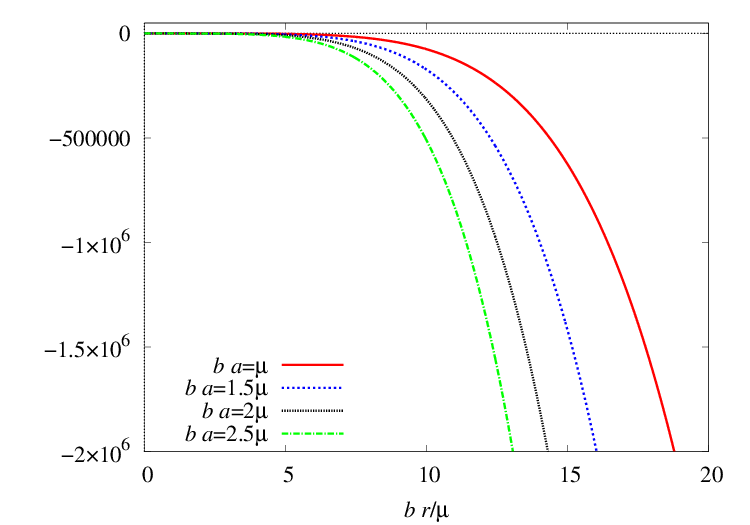}
    \caption{Validity of the $DEC_2$ considering different values of the parameter $a$.}
    \label{Fig-DEC2}
\end{figure}

For $DEC_2$, we need to check the validity of $\rho - p_{t} \geq 0$, which are written by:
\begin{equation}
\rho-p_t =\frac{a^2 \left(a^2-\tilde{r}^2\right) \left(b^2 \tilde{r}^2-3 b \mu  \tilde{r}+4 \mu ^2\right)}{2 \tilde{r}^8}\geq 0.
\end{equation}
For large $\tilde{r}$ and any parameters, we see that it is always violated. This can also be seen in Fig. \ref{Fig-DEC2}. As we can see, the value of the parameter $a$ does not significantly affect the result and the combination is always negative. Thus, regardless of the results for $WEC_2$, the $DEC_2$ condition will always be violated.

\begin{figure}
    \centering
    \includegraphics[scale=0.65]{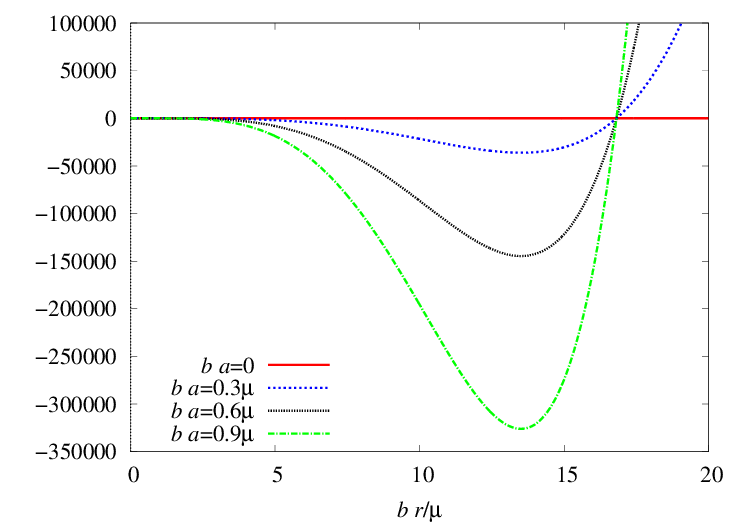}
    \includegraphics[scale=0.65]{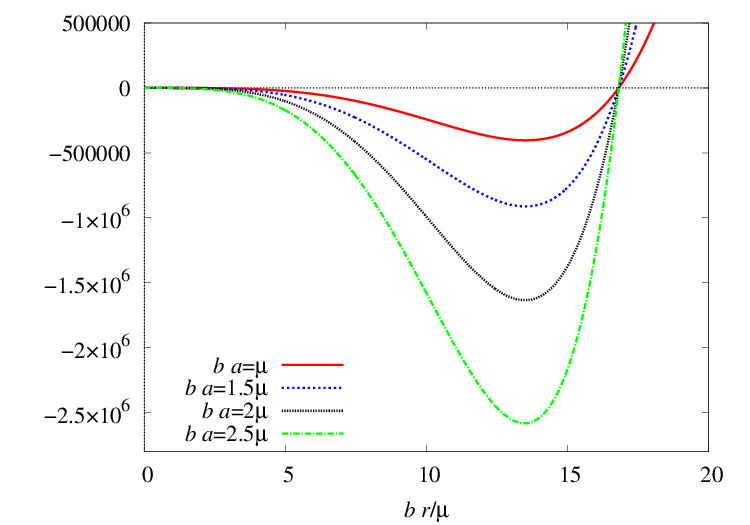}
    \caption{Validity of the $DEC_1$ considering different values of the parameter $a$.}
    \label{Fig-DEC1}
\end{figure}

For $DEC_1$, we need to check the validity of $\rho - p_{r} \geq 0$, which are written by:
\begin{equation}
    \rho-p_r=\frac{a^2b^2}{16 \tilde{r}^8} N_4(\tilde{r}) \geq 0\;; N_4(\tilde{r})=
8 \tilde{r}^4
- \frac{144 \mu}{b} \tilde{r}^3
+ \left( -14 a^2 + \frac{160 \mu^2}{b^2} \right) \tilde{r}^2
+ \frac{184 a^2 \mu}{b} \tilde{r}
- \frac{192 a^2 \mu^2}{b^2}.
\end{equation}
We see that for large $\tilde{r}$ it is satisfied and 
\begin{equation}
 N_4(a) =  -6 a^2\left(a^2-\frac{20 a \mu }{3 b}+\frac{16 \mu ^2}{3 b^2}\right).
\end{equation}
Therefore, at this point, the condition $\rho - p_{r}$ is satisfied in the range  
\begin{equation}
\frac{2 \mu}{3 b} \left( 5 - \sqrt{13} \right)
< a <
\frac{2 \mu}{3 b} \left( 5 + \sqrt{13} \right).
\end{equation}
The second derivative of $N_4$ is given by 
\begin{equation}
N''_4(\tilde{r})= 96 \tilde{r}^2 
- \frac{864 \mu}{b} \tilde{r} 
+ 2 \left( -14 a^2 + \frac{160 \mu^2}{b^2} \right).
\end{equation}
The interesting thing about the above expression is that its roots are given by 
\begin{equation}
\tilde{r}_{\pm} = \frac{1}{12 b} \left[ 54 \mu \pm \sqrt{42} \sqrt{a^2 b^2 + 58 \mu^2} \right].
\end{equation}
From the above roots, we can directly find $N_4(\tilde{r}_+)>0$. Following the same reasoning as before, we do not get a new condition to conclude about $N_4(\tilde{r})$. In Fig. \ref{Fig-DEC1}, we plot the cases where it is positive at the throat and can see that, even so, it will be negative for some regions.  However, we must also check the behavior of $WEC_1$, which is violated in this region, for any range of parameters. Therefore, we conclude that $DEC_1$ is always violated, for any chosen parameters.

For the case of a singular black hole, i.e., $a \to 0$, we have:
\begin{equation}
\rho=0, \quad \rho+p_r=0, \quad \rho+p_t=0, \quad \rho+p_r+p_t=0, \quad \rho-p_r=0, \quad \rho - p_t=0.    
\end{equation}
This shows that for the case of a singular black hole, the energy conditions are identically satisfied.

In this section, we examined the energy conditions of the regularized asymptotically locally flat solution in the context of purely quadratic gravity. In summary, we have shown that up to the $WEC_2$, all conditions are violated in some region of spacetime, irrespective of the range of parameters. For $WEC_2$, we were unable to draw such a strong conclusion; however, we found that for some range of parameters, it is satisfied for large $r$ and in the throat. By plotting it for some values in this range, we found its violation. Overall, the analysis emphasizes that the regularized solution requires exotic matter to meet certain energy conditions, especially in the context of WHs. This exotic matter is crucial for supporting such spacetimes, as indicated by the conditions being satisfied only under specific circumstances or parameter ranges.

\section{Geodesics}\label{S:Geo}
In order to study the trajectories of particles in the regularized spacetime, we can rewrite the line element of our regularized solution as
\begin{equation}
    ds^2=A(r)dt^2-C(r)dr^2-D(r)d\varphi ^2.\label{geoline}
\end{equation}

The trajectories of particles in this spacetime can be obtained through the Euler-Lagrange equations, where the Lagrangian associated with the line element is given by \cite{Chandrasekhar:1985kt}:
\begin{equation}
    2\mathcal{L}=\dot{s}^2=A\dot{t}^2-C\dot{r}^2-D\dot{\varphi}^2,\label{geoLagran}
\end{equation}
where the dot represents the derivative with respect to the affine parameter $\lambda$. The equations of motion are
\begin{eqnarray}
    A\dot{t}=E,\label{geo1}\\
   A\dot{t}^2-C\dot{r}^2-D\dot{\varphi}^2=\delta,\label{geo2}\\
   D\dot{\varphi}=\ell,\label{geo4}
\end{eqnarray}
where $E$ and $\ell$ are constants of motion that represent the energy and the angular momentum of the particle \cite{Chandrasekhar:1985kt}. For massive particles, we have $\delta=1$, and $\delta=0$ for massless particles. 
\subsection{Massless particles}
For massless particles, we can consider the initially presented line element, equation \eqref{line_reg}. Using the equations of motion to $A(r)=C(r)^{-1}=f(r)=b\sqrt{r^2+a^2}-\mu$ and $D(r)=\Sigma^2(r)=r^2+a^2$, we obtain
\begin{equation}
E^2=\dot{r}^2+\frac{\left(b\sqrt{r^2+a^2}-\mu\right)\ell^2}{r^2+a^2}.
\end{equation}
This is a type of energy conservation equation. It is important to note that $\dot{r}$ represents the radial velocity of the particle. We may rewrite this equation as
\begin{equation}
    \dot{r}^2=E^2-V_{eff}.\label{EnerConsMassless}
\end{equation}
The effective potential to massless particles is given by
\begin{equation}
V_{eff}= \frac{l^2 \left(b \sqrt{a^2+r^2}-\mu
   \right)}{a^2+r^2}.\label{Veffmassless}
\end{equation}

Through the first and second derivatives, we can assess whether this potential allows for the existence of stable or unstable orbits. From the condition $V'(r)=0$, we obtain that the potential has three extrema, which are given by
\begin{equation}
    r_{1}= \frac{\sqrt{4 \mu ^2-a^2 b^2}}{b}, \quad r_{2}=- \frac{\sqrt{4 \mu ^2-a^2 b^2}}{b},\quad \mbox{and} \quad r_{3}=0.\label{extremeSV}
\end{equation}
If $a < 2 \mu/ b$, we will have three possible orbits. For $a \geq 2\mu /b$, we have only one orbit at $r=0$.

\begin{figure}
    \centering
    \includegraphics[scale=0.65]{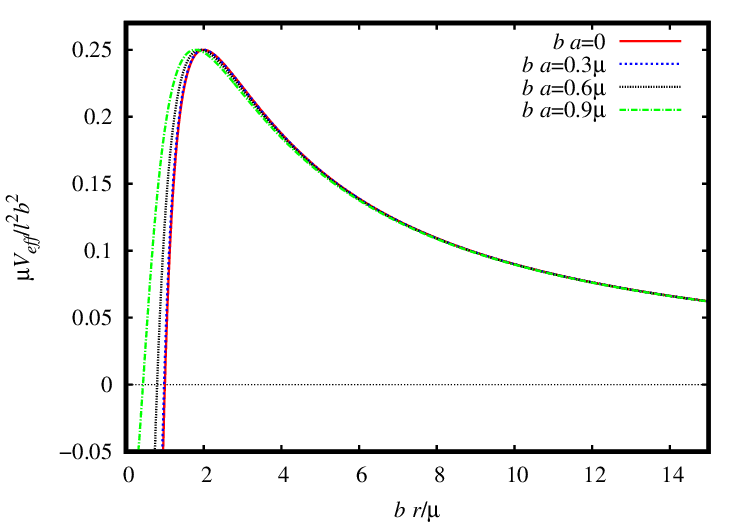}
    \includegraphics[scale=0.65]{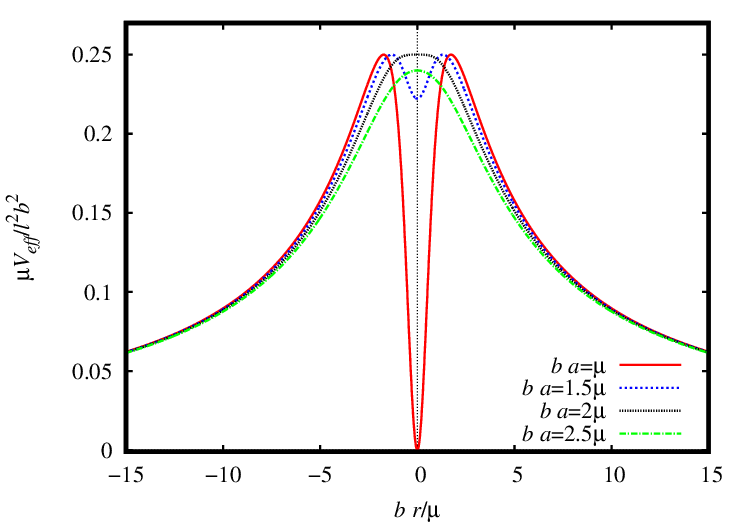}
    \caption{Effective potential for massless particles considering different values of the parameter $a$.}
    \label{Fig-Veff-SV-Massless}
\end{figure}

In Fig. \ref{Fig-Veff-SV-Massless}, we show the effective potential for massless particles. If $b a<\mu$, the maximum is located at $r=r_{1}$. If $\mu\leq ba < 2\mu$, the maximums are located at $r=r_{1}$ and $r=r_{2}$, and the minimum at $r=r_{3}$. If $b a\geq 2\mu$, there is only one unstable orbit at $r=r_{3}$. 

We can also calculate the radial acceleration, that is
\begin{equation}
  \ddot{r}=-\frac{1}{2}\frac{dV_{eff}}{dr}=\frac{\ell^2 r \left(a^2 b-2 \mu  \sqrt{a^2+r^2}+b
   r^2\right)}{2\left(a^2+r^2\right)^{5/2}}.\label{a_SV_massless}
\end{equation}
If we calculate the radial acceleration in the extrema points of the effective potential, we find $\ddot{r}(r_1,r_2,r_3)=0$. At these points, the radial velocity of the particle is also zero. This means that the particle follows a circular orbit around the black hole.

If we consider that the particle has an energy $E_c$, which is equal to the maximum of the effective potential, and using the equations of motion, we can obtain the angular velocity of the particle, given by 
\begin{equation}
    \dot{\varphi}^2=\pm \frac{E_c}{f(r_{1,2,3})^{1/2}\Sigma(r_{1,2,3})},
\end{equation}
which is nonzero. Since the angular velocity is related to the angular momentum $\ell$, we can use this to obtain the expression for the critical impact parameter, which is given by \cite{Chandrasekhar:1985kt}
\begin{equation}
    B_c=\frac{\ell_c}{E_c}=(r_c^2+a^2)^{1/2}\left(b \sqrt{r_c^2+a^2}-\mu\right)^{-1/2},
\end{equation}
where $E_c$ and $\ell_c$ are the total energy and angular momentum of the particle in a circular orbit, and $r_c$ is the radius of the circular orbit that the particle describes. Here, $r_c=r_1$ to $0\leq b a< 2\mu$ and $r_c=r_3$ if $b a\geq 2\mu$. Substituting $r_1$ and $r_3$ we find
\begin{eqnarray}
    B_c&=&\frac{2 \sqrt{\mu }}{b}, \quad \mbox{if} \quad ba< 2\mu,\\
    B_c&=&\frac{a}{\sqrt{a b-\mu }}, \quad \mbox{if} \quad b a\geq 2\mu.
\end{eqnarray}
The impact parameter will tell us what will happen to the particle. If $B > B_c$, the particle will be scattered; if $B < B_c$, the particle will be absorbed; and for $B = B_c$, the particle will perform circular motion around the black hole. In $3+1$ dimensions, we have the absorption cross-section. In the case of $2+1$ dimensions, we have the absorption length, which is defined as $\sigma = 2B_c$ \cite{Oliveira:2010zzb}. Any particle emitted from infinity and within this length interval will be absorbed by the black hole.
For the spacetime we are considering, the absorption length is given by
\begin{eqnarray}
    \sigma&=&\frac{4 \sqrt{\mu }}{b}, \quad \mbox{if} \quad ba< 2\mu,\\
    \sigma&=&\frac{2a}{\sqrt{a b-\mu }}, \quad \mbox{if} \quad b a\geq 2\mu.
\end{eqnarray}
For $ba<2\mu$, we have the same absorption length as in the singular case, which does not depend explicitly on $a$. For $ba\geq2\mu$, the absorption length will grow with $a$.

\begin{figure}
    \centering
   \subfigure[]{\includegraphics[scale=0.8]{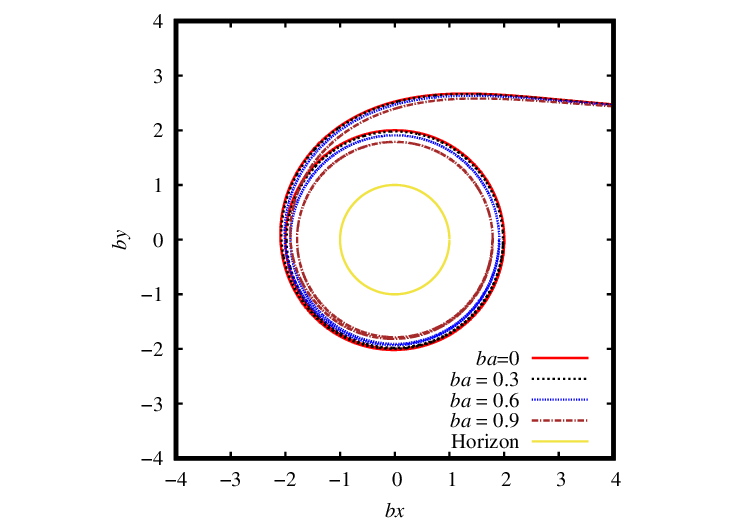}}\hspace{-3cm}
    \subfigure[]{\includegraphics[scale=0.8]{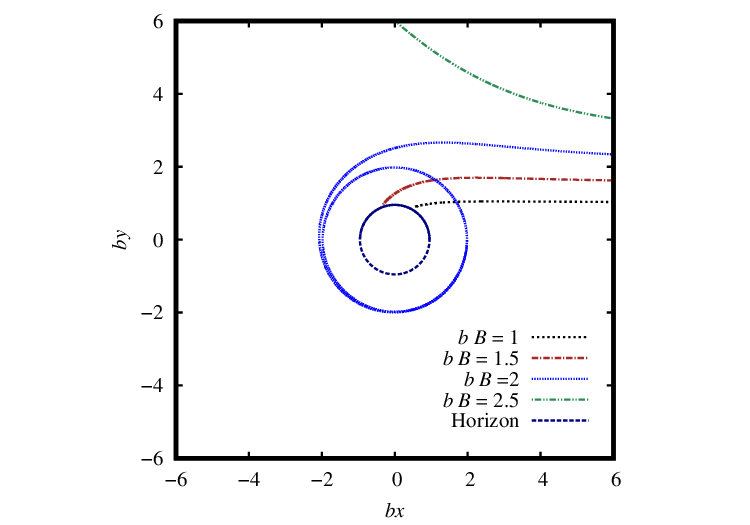}}
    \caption{Trajectories of massless particles considering $\mu=1$. In (a) we fixed the value of the impact parameter, $bB=bB_c=2$, and varied the value of $a$. We place the event horizon only for the case of the largest radius, a = 0. The other values have horizons with smaller radii, and therefore, we choose not to insert them. In (b) we fix the value $ba=0.3$ and change the value of the impact parameter.}
    \label{fig:SV-geo-massless}
\end{figure}

Through equation \eqref{EnerConsMassless} and making the change of variables $u = 1/\sqrt{r^2 + a^2}$, we obtain
\begin{equation}
    \left(\frac{du}{d\varphi}\right)^2=\left(1-a^2u^2\right)\left(\frac{1}{B^2}+u (\mu  u-b)\right).
\end{equation}
Taking the derivation of the equation above with respect to $\varphi$ we find
\begin{equation}
    \frac{d^2u}{d\varphi^2}-\frac{3}{2} a^2 b u^2+\frac{a^2 u}{B^2}+2 a^2 \mu  u^3+\frac{b}{2}-\mu  u=0.\label{EQ2Geo-SV-massless}
\end{equation}
To solve the equation \eqref{EQ2Geo-SV-massless}, we need two boundary conditions that are
\begin{equation}
\left.u\right|_{\varphi=0}=0, \quad \mbox{and} \quad    \left.\frac{du}{d\varphi}\right|_{\varphi =0}=\frac{1}{B}.
\end{equation}
In Fig. \ref{fig:SV-geo-massless}, for the case in which we vary the parameter $a$ while fixing the critical impact parameter, the radius of the unstable orbit decreases as the parameter $a$ increases. This also occurs in solutions such as Reissner-Nordstrom. In the case where the impact parameter is varied, there will be values where the particle is absorbed by the black hole, scattered, or remains in an unstable circular orbit. It is interesting to note that, in the scattering case, the particle undergoes very intense scattering.

\subsection{Geodesics of massive particles}
For massive particles, we consider the equations of motion with $\delta = 1$,
\begin{equation}
    f(r)\dot{t}^2-f(r)^{-1}\dot{r}^2-\Sigma(r)^2\dot{\varphi}^2=1.
\end{equation}
Using \eqref{geo1}-\eqref{geo4}, we find
\begin{equation}
   E^2=\dot{r}^2+V_{eff}.\label{EnerConsMassive}
\end{equation}
The structure of the equation above is similar to the massless case. The differences are incorporated within the effective potential, which is given by: 
\begin{equation}
   V_{eff}=f\left(1+\frac{\ell^2}{\Sigma^2}\right)=\left(b\sqrt{r^2+a^2}-\mu\right)\left(1+\frac{\ell^2}{a^2+r^2}\right).
\end{equation}
\begin{figure}
   \centering
  \subfigure[]{\includegraphics[scale=0.65]{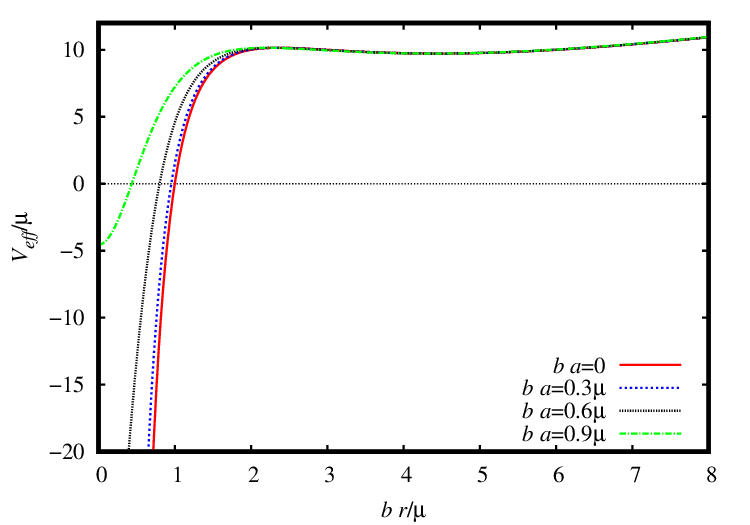}}
  \subfigure[]{\includegraphics[scale=0.65]{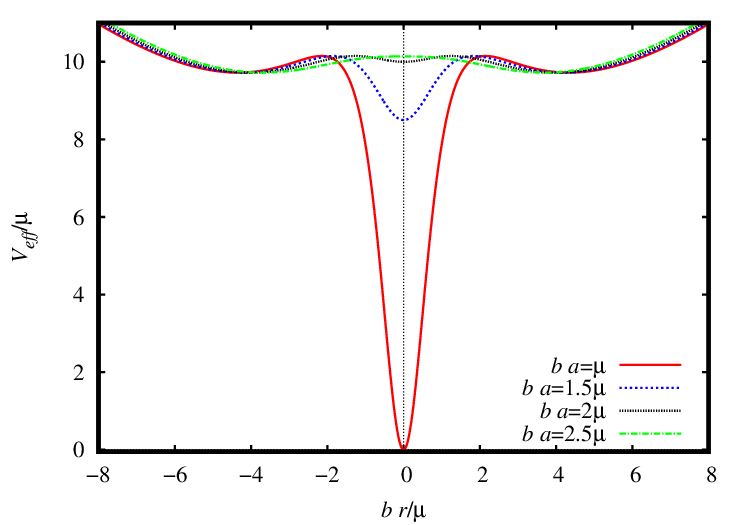}}
  \subfigure[]{\includegraphics[scale=0.65]{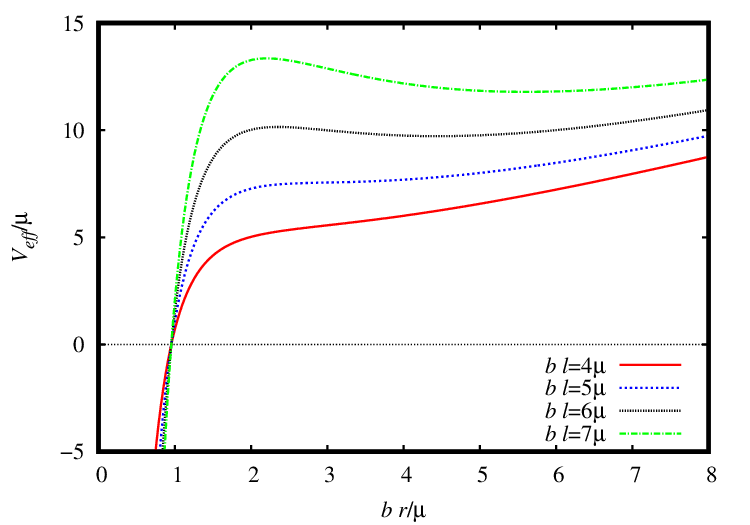}}
  \subfigure[]{\includegraphics[scale=0.65]{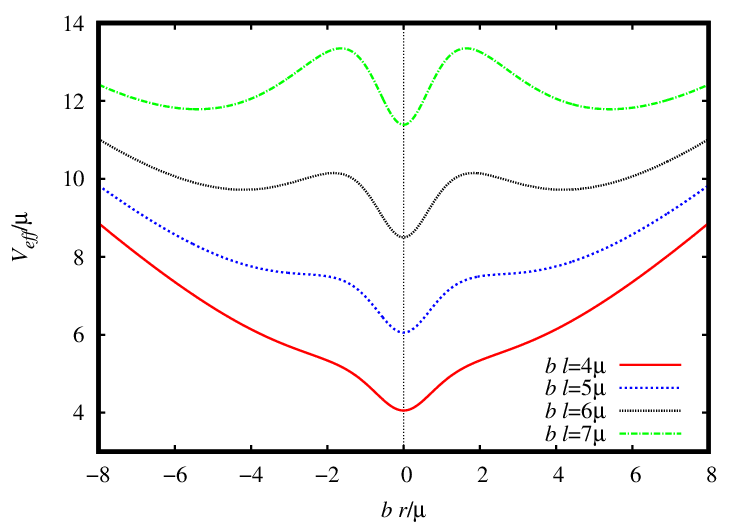}}
    \caption{Effective potential to massive particles to different values of $a$ and $\ell$. In (a) we fix $b\ell=6$ and change the $a$. As to $ba<m$ there is an event horizon,  we consider only the region $r>0$. In (b) the configuration is similar of (a), however in this case there is no event horizon and we consider $-\infty<r<\infty$. In (c) and (d) we fix the charge, $ba=0.3\mu$ and $ba=1.5\mu$ respectively, and change $\ell$.}
    \label{Fig-Veff-SV-Massive}
\end{figure}

In Fig. \ref{Fig-Veff-SV-Massive} we show the behavior of the effective potential for different values of $a$ and $\ell$. Both the parameters $a$ and $\ell$ can alter the structure of the effective potential. For small values of $a$, there are two extrema in the potential, one maximum and one minimum. This means that we can have both stable and unstable orbits outside the event horizon. For $ba>\mu$, we no longer have an event horizon, and as a result, the orbit at $r=0$ also becomes accessible, potentially being either a maximum or minimum point, depending on the value of $a$. The change in angular momentum of the particle does not affect the presence of the event horizon but does alter the structure of the effective potential. In the case with horizons, particles with low angular momentum are necessarily absorbed by the black hole, whereas particles with high angular momentum can remain in a stable or unstable circular orbit.  

To obtain the extremal points, we must solve $V_{eff}'=0$. However, the analytical expressions are quite complex, so we will analyze them graphically. 

\begin{figure}
    \centering
    \subfigure[]{\includegraphics[scale=0.65]{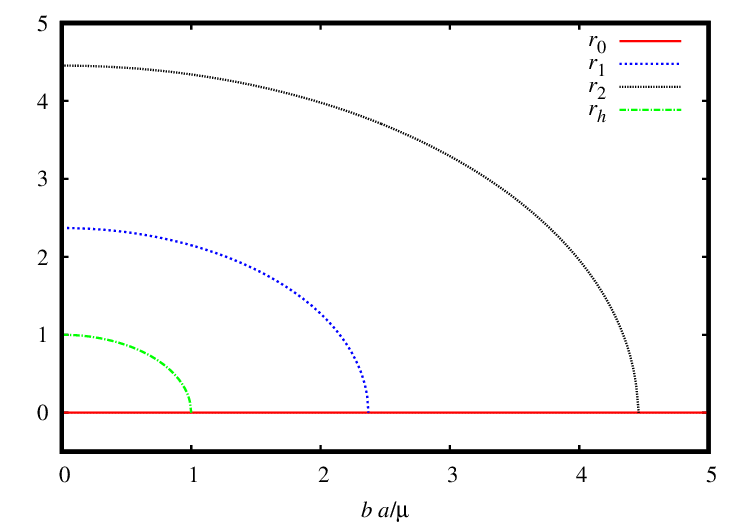}}
    \subfigure[]{\includegraphics[scale=0.65]{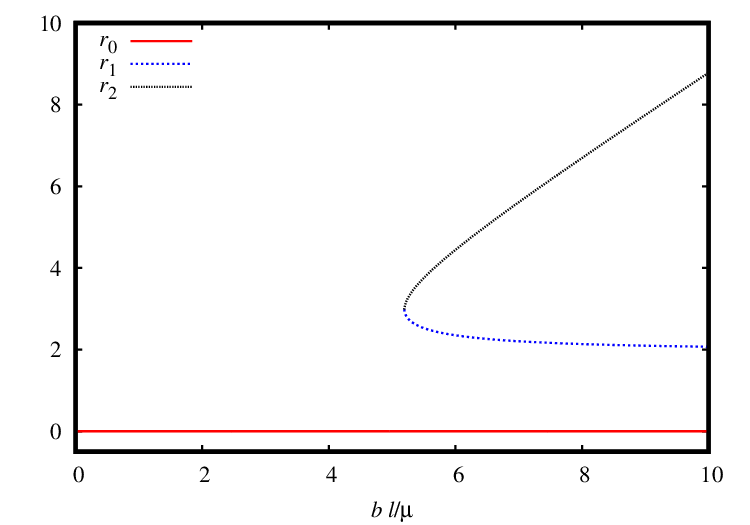}}
    \caption{Radius of the stable and the unstable orbit to massive particles as functions of the $a$ and $\ell$. In (a), we fix the value of angular momentum at $b\ell=6$ and vary the value of $a$. In Figure (b), we fix $ba=0.3\mu$ and vary the value of angular momentum. }
    \label{fig:SV-radiusmassive}
\end{figure}

\begin{figure}
    \centering
    \includegraphics[scale=0.65]{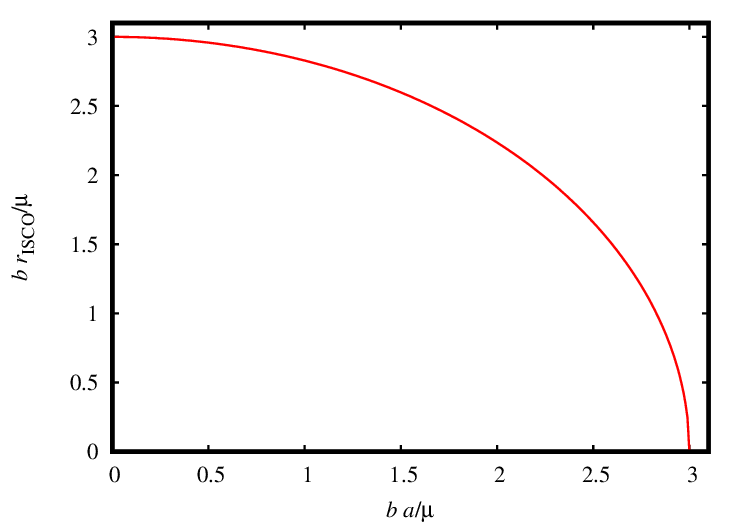}
   \caption{Radius of ISCO as functions of the parameter $a$.}
    \label{fig:SV-ISCO}
\end{figure}

In Fig. \ref{fig:SV-radiusmassive}, we analyze the radius of the circular orbits. We observe that there are three possible orbits. However, as long as there is an event horizon, only two orbits are accessible, one stable and one unstable. As the $a$ parameter increases, the horizon ceases to exist and the orbit at $r_0=0$ also becomes possible. Since the solution is symmetric for $r\rightarrow -r$, the orbits also exist for negative $r$. For great values of $a$, only orbits at $r=0$ exist. As we change the angular momentum of the particle, the radius of the orbits are also altered. There is a value of $\ell$ such that the stable and unstable orbits tend to the same point, this is the innermost stable circular orbit (ISCO). For values of $\ell$ smaller than $l_{ISCO}=3 \sqrt{3} \mu/ b$, only the orbit at $r=0$ exists if there is no event horizon.  For $\ell=\ell_{ISCO}$, we find the value of the ISCO radius, which is given by 
\begin{equation}
   r_{ISCO}=\sqrt{\frac{9 \mu ^2}{b^2}-a^2}.
\end{equation}
In Fig. \ref{fig:SV-ISCO},we analyze the change in the ISCO radius for different values of the regularization parameter. If $a=0$ we have $r_{ISCO}=3\mu/b$, and if $a=3\mu/b$ we have $r_{ISCO}\rightarrow r_0=0$.

From equation \eqref{EnerConsMassive}, we can define the following function $G(r)$:
\begin{equation}
    G(r)=\left(\frac{dr}{d\phi}\right)^2=\frac{\left(a^2+r^2\right)^2 }{\ell^2}\left(E^2-\left(\frac{\ell^2}{a^2+r^2}+1\right) \left(b
   \sqrt{a^2+r^2}-\mu \right)\right).
\end{equation}
The zeros of this function represent the particle's return points. 

\begin{figure}
    \centering
    \includegraphics[scale=0.65]{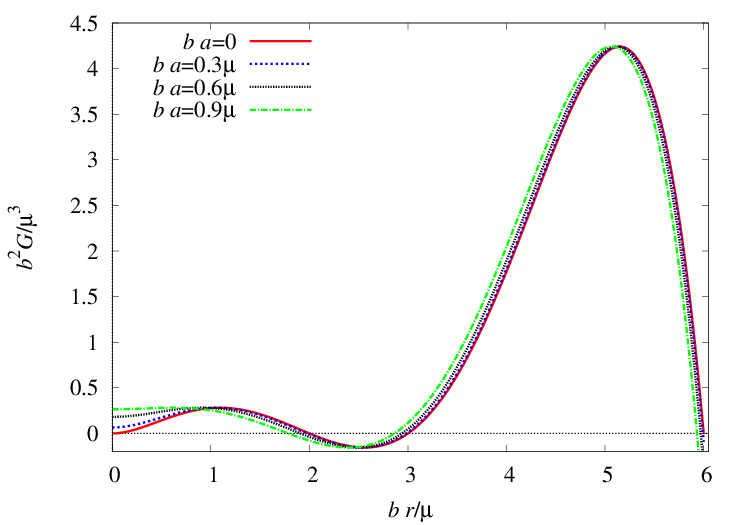}
    \includegraphics[scale=0.65]{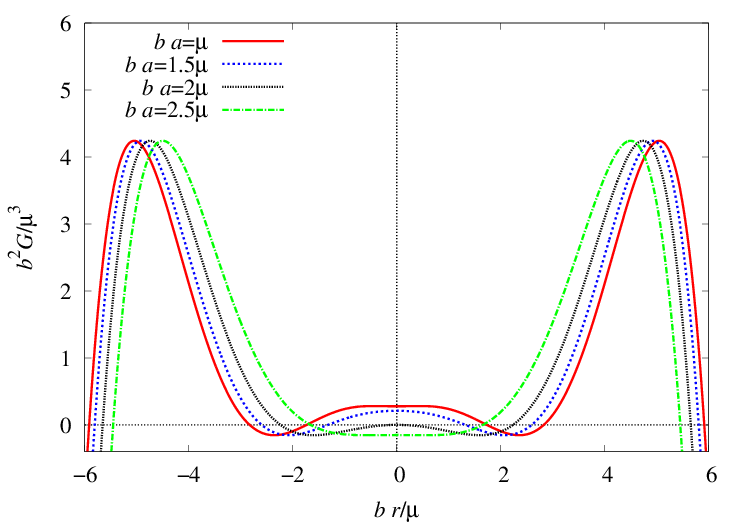}
    \includegraphics[scale=0.65]{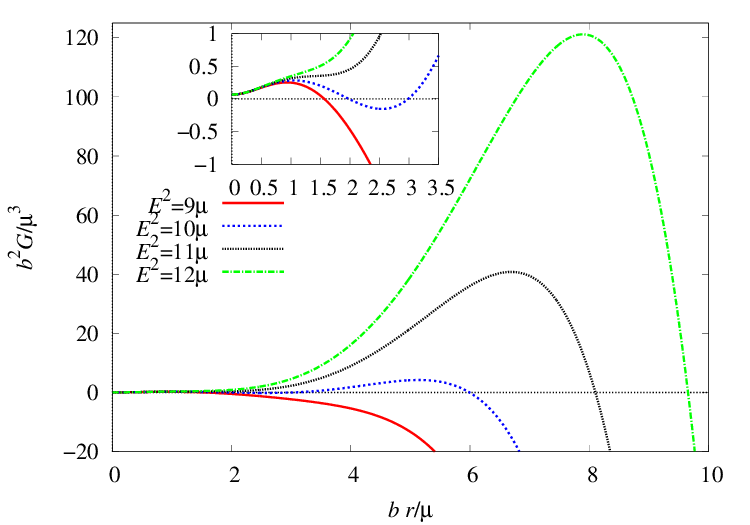}
    \includegraphics[scale=0.65]{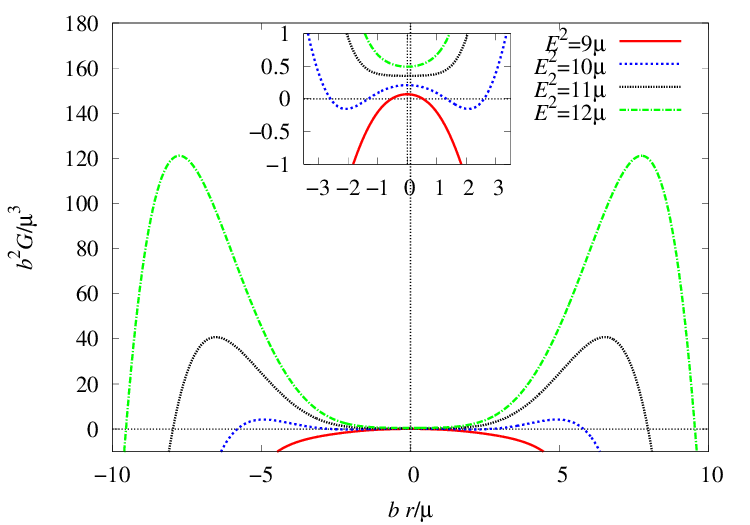}
    \caption{Behavior of the function $G$ for massive particles as a function of the radial coordinate for different values of the particle's energy and the parameter $a$. In all panels, the particle's angular momentum is fixed at $b\ell=6\mu$. In panels (a) and (b), we fix the energy value at $E^2=10\mu$ and vary the parameter $a$. In panel (a), cases of a RBH are represented, as the event horizon still exists. In panel (b), we have the case of a black throat and a WH. In panels (c) and (d), we fix $ba=0.3\mu$ and $ba=1.5\mu$, respectively, and vary the value of the particle's energy.}
    \label{fig:Gmassive}
\end{figure}

Figure \ref{fig:Gmassive} illustrates the behavior of the function $G(r)$ in different cases. In the case with energy $E^2=10\mu$, for some values of the parameter $a$, every particle emitted from distant points of the black hole reaches a maximum approach point and then returns. The higher the value of $a$, the closer the particle can reach the black hole or throat. As expected from the shape of the effective potential, massive particles cannot be emitted from infinity, once the radial velocity must be real and not imaginary. Depending on the energy value, the particle may reach a return point before reaching the black hole/WH or it may reach it. The higher the particle's energy, the farther it can be emitted. For a particle to be emitted from infinity, it would need to have infinite energy. In the case of a WH, a particle emitted from distant points can only traverse it if it has a minimum energy value; otherwise, it will encounter a return point.

\section{Conclusions and discussion}\label{S:conclusions}
In this work, we apply the Simpson-Visser regularization method to an asymptotically locally flat solution in $2+1$ dimensions in the context of purely quadratic gravity, $K$ gravity. Once we have the regularized solution, we analyze properties and extract information from this regularized spacetime. 

The regularized solution has a minimum circumference $\mathcal{C}=2\pi a$, and thus we have a kind of WH hidden inside a black hole, which characterizes a BB. The BB will only be present while $0<a<\mu/b$. If the regularization parameter is null, we recover the singular solution, and if $a = \mu/b$, the WH's throat coincides with the event horizon, and, with this, we have a black throat, which is the throat of a WH that is traversable in one direction. If $a > \mu/b$, there will no longer be an event horizon and, with this, we will have a WH that is traversable in both directions. 
For the case with a negative cosmological constant, the causal structure becomes richer by allowing the presence of event horizons even in cases where the mass or the parameter $b$ are negative. This brings more possibilities for the causal structure of spacetime.

We were able to identify the source fields that generate the regular solution in quadratic theory. The source is described by a combination of a partially phantom scalar field along with a NED. The analysis of the nature of the scalar field is conducted through the function $h(\phi)$, which accompanies the kinetic term of the scalar field. We found that the function $h(\phi)$ has positive values near the center of the solution and negative values at more distant points. This implies that our scalar field will be canonical in regions close to the throat and phantom in regions further outside the WH throat.

The analysis of the energy conditions for the regularized asymptotically locally flat solution reveals notable behaviors based on the interplay between the parameters $a$, $b$, and $\mu$. The inequalities related to the dominant energy condition will always be violated, whereas the other inequalities may be partially satisfied in certain regions of spacetime. The energy conditions will only be satisfied in all regions of spacetime when considering the singular case, as we have vacuum, and, therefore, the inequalities are identically satisfied.
This implies that the presence of exotic matter, which is crucial for supporting such spacetimes, is determined by the fluid quantities and their relationships, especially in the context of WHs. This indicates that the problem of exotic matter persists for both RBHs and WHs in $K$-gravity theories.

Through the analysis of geodesics, we find that there is only one unstable orbit for massless particles in cases where there is an event horizon and two unstable orbits and one stable for some cases of WHs. For cases with an event horizon, the critical impact parameter does not depend on the regularization parameter, while the radius of the orbit of massless particles depends on this parameter. In fact, the radius of the unstable orbit decreases as the value of the regularization parameter increases. There are some values of the impact parameter where the particles are strongly scattered. Like the critical impact parameter, the absorption length, in cases where event horizons are present, does not depend on the regularization parameter. For cases where the absorption length depends on the regularization parameter, the length will increase as the value of $a$ increases.

For massive particles, it is possible to have stable and unstable orbits, both for the case with event horizons and for the WH case. As the regularization parameter increases, the event horizon ceases to exist and another stable orbit arises. If the particles have sufficient angular momentum, it is possible to have both stable and unstable orbits. However, for particles with low angular momentum, there will only be stable orbits and this only in the case without horizons.

Although BB models in $2+1$ dimensions already exist in the literature \cite{Furtado:2022tnb,Alencar:2024yvh}, our work presents the first asymptotically flat solution in this context, which significantly distinguishes it from the previously regularized versions of the BTZ black hole. Furthermore, we are the first to explicitly identify the matter sources that give rise to black bounce solutions in alternative theories of gravity, an important step forward, even though other studies have later emerged in this direction. Another distinctive feature of our model is the presence of a partially phantom scalar field among the sources, in contrast to most existing solutions where the scalar field is fully phantom. Finally, we show that the null energy condition is not necessarily violated throughout the entire spacetime, which represents yet another novel aspect of our approach.

Finally, we point out that this is a series of papers where we will apply the method used here to find sources for alternative theories of gravity. 

\section*{Acknowledgments}
We thank Professor Kirill Bronnikov for his comments and suggestions. G.A. and M.E.R. would like to thank Conselho Nacional de Desenvolvimento Cient\'ifico e 
Tecnol\'ogico - CNPq, Brazil  for partial financial support. G.A. and M.S. would like to thank 
Funda\c c\~ao Cearense de Apoio ao Desenvolvimento Cient\'ifico e Tecnol\'ogico (FUNCAP) 
for partial financial support.


\bibliography{ref.bib}
\end{document}